\documentclass[]{article} 
\usepackage{latexsym} 
\usepackage{amssymb}
\usepackage{amsmath} 
\usepackage{amsthm} 
\usepackage{color}
\usepackage{graphics}
\usepackage{graphicx}
\usepackage{epsfig}

\newtheorem{theorem}{Theorem}
\newtheorem*{theorem*}{Theorem}

\newtheorem{lemma}{Lemma}
\newtheorem{corollary}{Corollary}

\theoremstyle{definition}
\newtheorem{definition}{Definition}
\newtheorem{notation}{Notation}

\newtheorem{remark}{Remark}

\newcommand{\fA}{\ensuremath{\mathfrak{A}}} 

\newcommand{\cC}{\ensuremath{\mathcal{C}}}  
\newcommand{\cF}{\ensuremath{\mathcal{F}}}  
\newcommand{\cI}{\ensuremath{\mathcal{I}}}  

\newcommand{\bbN}{\ensuremath{\mathbb{N}}}   

\title{Complexity of the Two-Variable Fragment with (Binary Coded)
  Counting Quantifiers}
\author{Ian Pratt-Hartmann\\
  Department of Computer Science,\\
  University of Manchester.}  \date{}
\begin{document}
\maketitle
\begin{abstract}
\noindent
We show that the satisfiability and finite satisfiability problems for
the two-variable fragment of first-order logic with counting
quantifiers are both in NEXPTIME, even when counting quantifiers are
coded succinctly.
\end{abstract}

\section{Background}
\label{sec:background}
The {two-variable fragment with counting quantifiers}, here denoted
$\cC^2$, is the set of function-free, first-order formulas containing
at most two variables, but with the counting quantifiers
$\exists_{\leq C}$, $\exists_{\geq C}$ and $\exists_{= C}$ (for every
$C >0$) allowed.  The {\em satisfiability problem}, Sat-$\cC^2$, is
the problem of deciding, for a given sentence $\phi$ of $\cC^2$,
whether $\phi$ has a model; the {\em finite satisfiability problem},
Fin-Sat-$\cC^2$, is the problem of deciding, for a given sentence
$\phi$ of $\cC^2$, whether $\phi$ has a finite model. It is well-known
that $\cC^2$ lacks the finite model property; hence Sat-$\cC^2$ and
Fin-Sat-$\cC^2$ do not coincide. The decidability of Sat-$\cC^2$ was
shown independently by Gr\"{a}del, Otto and Rosen~\cite{logic:gor97},
and by Szwast and Tendera~\cite{logic:pst97,logic:pst99}. For a more
general survey and some recent extensions, see Gr\"{a}del and
Otto~\cite{logic:GO99} and Otto~\cite{logic:otto01}.

When discussing the computational complexity of these problems, it is
important to specify how the sizes of numerical quantifier subscripts
are measured. Under {\em unary coding}, a counting subscript $C$ is
taken to have size $C$; under {\em binary coding}, by contrast, the
same subscript is taken to have size $\log C$.  In determining upper
complexity-bounds, unary coding is evidently the less stringent
accounting method, because the size of formulas involving counting
quantifiers is, in general, exponentially larger than under binary
coding. Pacholski {\em et al., op.~cit.} showed that Sat-$\cC^2$ is in
NEXPTIME, but only under the assumption of unary coding. The present
paper shows that both Fin-Sat-$\cC^2$ and Sat-$\cC^2$ are in NEXPTIME,
even with binary coding.  We remark that both Fin-Sat-$\cC^2$ and
Sat-$\cC^2$ are easily shown to be NEXPTIME-hard on either coding
scheme.

In the sequel, we confine attention to finite or countably infinite
structures interpreting finite signatures of unary and binary
predicates, without individual constants or function-symbols; in
addition, we treat the equality predicate $\approx$ as a logical
constant. The lack of individual constants in the signature
constitutes no essential restriction of expressive power, because
their effect can be reproduced using formulas of the form
$\exists_{=1} x p(x)$. The presence of equality in the logic
constitutes no extension of expressive power, because it can be
defined by the formula $\forall x (x \approx x) \wedge \forall x
\exists_{=1} y (x \approx y)$.

\section{Preliminaries}
\label{sec:preliminaries}
If $\phi$ is a formula, let $\lVert \phi \rVert$ denote the number of
symbols in $\phi$, assuming that a numerical quantifier subscript $C$
involves $\log C$ symbols. Thus, if the quantifier $\exists_{=C}$
occurs in $\phi$, we have $C \leq 2^{\lVert \phi \rVert}$. The
following type of normal-form lemma is standard in treatments of
$\cC^2$. (Recall that, in this paper, we are confining attention to
signatures with no individual constants or function-symbols.)
\begin{lemma}
Let $\phi$ be a sentence in $\cC^2$. We can construct, in time bounded
by a polynomial function of $\lVert \phi \rVert$, a formula
\begin{equation*}
\phi^* := 
\forall x \alpha \wedge 
 \forall x \forall y (\beta \vee x \approx y) \wedge 
   \bigwedge_{1 \leq h \leq m} 
      \forall x \exists_{=C_h} y (f_h(x,y) \wedge x \not \approx y)
\end{equation*}
satisfying the following conditions: $(i)$ $\alpha$ is a
quantifier-free, equality-free formula with $x$ as its only variable,
$(ii)$ $\beta$ is a quantifier-free, equality-free formula with $x$
and $y$ as its only variables, $(iii)$ the $f_h$ $(1 \leq h \leq m)$
are binary predicates, $(iv)$ $\phi^*$ is satisfiable if and only if
$\phi$ is satisfiable over a domain of size greater than $C = \max_{1
\leq h \leq m} C_h$.
\label{lma:normalform}
\end{lemma}
\begin{proof}
Routine adaptation of textbook transformation to Scott normal form.
See, for example, Gr\"{a}del and Otto~\cite{logic:GO99} Sec.~2.1 for
an explanation of the required techniques.
\end{proof}
We remark that, since $C$ is bounded by a singly exponential function
of $\lVert \phi \rVert$, satisfiability of $\phi$ over domains of size
$C$ or less can certainly be determined in NEXPTIME.

We next review some standard concepts.  Fix a finite signature
$\Sigma$ of unary and binary predicates (no individual constants
or function-symbols). A {\em literal} is an atomic
formula or the negation of an atomic formula.  A {\em 1-type} is a
maximal consistent set of (equality-free) literals involving only the
variable $x$. A {\em 2-type} is a maximal consistent set of
equality-free literals involving only the variables $x$ and $y$. If
$\tau$ is a 2-type, then the result of transposing the variables $x$
and $y$ in $\tau$ will also be a 2-type, denoted $\tau^{-1}$.  If
$\fA$ is any structure interpreting $\Sigma$, and $a \in A$, then
there exists a unique 1-type $\pi(x)$ such that $\fA \models \pi[a]$;
we denote $\pi$ by ${\rm tp}^\fA[a]$. If, in addition, $b \in A$ is
distinct from $a$, then there exists a unique 2-type $\tau(x,y)$ such
that $\fA \models \tau[a,b]$; we denote $\tau$ by ${\rm tp}^\fA[a,b]$.
We do not define ${\rm tp}^\fA[a,b]$ if $a = b$.  If $\pi$ is a
1-type, we say that $\pi$ is {\em realized} in $\fA$ if there exists
$a \in A$ with ${\rm tp}^\fA[a] = \pi$.  If $\tau$ is a 2-type, we say
that $\tau$ is {\em realized} in $\fA$ if there exist distinct $ a, b
\in A$ with ${\rm tp}^\fA[a,b] = \tau$.
\begin{notation} 
Any 2-type $\tau$ includes a unique 1-type, denoted ${\rm
tp}_1(\tau)$; in addition we write ${\rm tp}_2(\tau)$ for ${\rm
tp}_1(\tau^{-1})$.
\label{notation:types}
\end{notation}
\begin{remark}
If ${\rm tp}^\fA[a,b] = \tau$, then ${\rm tp}^\fA[b,a] = \tau^{-1}$,
${\rm tp}^\fA[a] = {\rm tp}_1(\tau)$ and ${\rm tp}^\fA[b] = {\rm
tp}_2(\tau)$.
\label{remark:types}
\end{remark}

We next introduce some non-standard concepts.
\begin{definition}
  A {\em counting signature} $\Sigma$ is a signature of unary and
  binary predicates (no individual constants or function-symbols) with
  a distinguished subset of binary predicates; we refer to these
  distinguished binary predicates as the {\em counting predicates} of
  $\Sigma$.
\label{def:countingsignature}
\end{definition}
Counting signatures help us analyse models of formulas in $\cC^2$ by
allowing us to treat a subset of the binary predicates in a special
way.  Since counting signatures are signatures, we may meaningfully
speak of 1-types and 2-types over counting signatures.
\begin{definition}
  Let $\Sigma$ be a counting signature, and let $\tau$ be a 2-type over
  $\Sigma$.  We say that $\tau$ is a {\em message-type} (over $\Sigma$)
  if, for some counting predicate $f$ of $\Sigma$, $f(x,y) \in \tau$.
  If $\tau$ is a message-type such that $\tau^{-1}$ is also a
  message-type, we say that $\tau$ is {\em invertible}. On the other
  hand, if $\tau$ is a 2-type such that neither $\tau$ nor $\tau^{-1}$
  is a message-type, we say that $\tau$ is {\em silent}.
\label{def:messagetype}
\end{definition}
Thus, a 2-type $\tau$ is an invertible message-type if and only if
there are counting predicates $f$ and $g$ such that $f(x,y) \in \tau$
and $g(y,x) \in \tau$.  The terminology is meant to suggest the
following imagery.  If ${\rm tp}^\fA[a,b]$ is a message-type $\mu$,
then we may imagine that $a$ sends a message (of type $\mu$) to $b$.
(On this view, $a$ can send at most one message to $b$). And if $\mu$
is invertible, then $b$ replies by sending a message (of type
$\mu^{-1}$) back to $a$. If ${\rm tp}^\fA[a,b]$ is silent, then
neither element sends a message to the other.

The remainder of this section is devoted to three simple observations
on structures interpreting counting signatures.
\begin{definition}
  Let $\fA$ be a structure interpreting a counting signature $\Sigma$,
  and let $Y$ be an integer. We say that $\fA$ is $Y$-{\em branching}
  ({\em over} $\Sigma$) if, for every $a \in A$ and every counting
  predicate $f$ of $\Sigma$, the set $\{a' \in A | a \neq a' \mbox{
  and } \fA \models f[a,a'] \}$ has cardinality at most $Y$. We say
  that $\fA$ is {\em finitely branching} if $\fA$ is $Y$-{\em
  branching} for some $Y \in \bbN$.
\label{def:branching}
\end{definition}
\begin{definition}
  Let $\fA$ be a structure interpreting a counting signature $\Sigma$.
  We say that $\fA$ is {\em chromatic} ({\em over} $\Sigma$) if, for
  all $a, a', a'' \in A$, the following two conditions hold:
\begin{enumerate}
\item if $a \neq a'$ and ${\rm tp}^\fA[a,a']$ is an invertible
  message-type, then ${\rm tp}^\fA[a] \neq {\rm tp}^\fA[a']$;
\item if $a, a', a''$ are pairwise distinct and both ${\rm
  tp}^\fA[a,a']$ and ${\rm tp}^\fA[a',a'']$ are invertible
  message-types, then ${\rm tp}^\fA[a] \neq {\rm tp}^\fA[a'']$.
\end{enumerate}
\label{def:chromatic}
\end{definition}
Thus, a chromatic structure is one in which distinct elements
connected by a chain of 1 or 2 invertible message-types have distinct
1-types.
\begin{lemma}
  Let $\Sigma$ be a counting signature with exactly $m$ counting
  predicates, let $Y$ be a non-negative integer, and let $\fA$ be a
  $Y$-branching structure interpreting $\Sigma$.  Let $\Sigma'$ be the
  signature formed by adding $\log((mY)^2 + 1)$ $($rounded up$)$ new
  unary predicates to $\Sigma$, and regard $\Sigma'$ as a counting
  signature by taking the counting predicates of $\Sigma'$ to be the
  same as the counting predicates of $\Sigma$.  Then $\fA$ can be
  expanded to a chromatic structure $\fA'$ interpreting $\Sigma'$.
\label{lma:chromatic}
\end{lemma}
\begin{proof}
  Consider the (undirected) graph $G$ on $A$ whose edges are the
  pairs of distinct elements connected by a chain of 1 or 2 invertible
  message-types. That is, $G = (A,E^1 \cup E^2)$, where
\begin{align*}
E^1 =& \{ (a,a') \mid  \mbox{$a \neq a'$ and ${\rm tp}^\fA[a,a']$ is an 
                           invertible message-type} \} \\
E^2 =& \{ (a,a'') \mid \mbox{$a \neq a''$ and for some $a' \in A$,
                            $(a,a')$ and $(a',a'')$ are both in $E^1$}\}.
\end{align*}
Since there are only $m$ counting predicates in $\Sigma$ and $\fA$ is
$Y$-branching, the degree of $G$ (in the normal graph-theoretic sense)
is at most $(mY)^2$. Now use the standard greedy algorithm to colour
$G$ with $(mY)^2 + 1$ colours.  By interpreting the $\log((mY)^2 + 1)$
(rounded up) new unary predicates to encode these colours, we obtain
the desired expansion $\fA'$.
\end{proof}
\begin{definition}
  Let $\fA$ be a structure and $Z$ an integer. We say that $\fA$ is
  $Z$-{\em differentiated} if, for every 1-type $\pi$, the number $u$
  of elements in $A$ having 1-type $\pi$ satisfies either $u \leq 1$
  or $u > Z$.
\label{def:differentiated}
\end{definition}
\begin{lemma}
  Let $\Sigma$ be a counting signature, let $\fA$ be a structure
  interpreting $\Sigma$, and let $Z$ be a positive integer. Let
  $\Sigma'$ be the signature formed by adding $\log Z$ $($rounded
  up$)$ new unary predicates to $\Sigma$, and regard $\Sigma'$ as a
  counting signature by taking the counting predicates of $\Sigma'$ to
  be the same as the counting predicates of $\Sigma$.  Then $\fA$ can
  be expanded to a $Z$-differentiated structure $\fA'$ interpreting
  $\Sigma'$. Moreover, if $\fA$ is chromatic over $\Sigma$, we can
  ensure that $\fA'$ is chromatic over $\Sigma'$.
\label{lma:differentiated}
\end{lemma}
\begin{proof}
  For each 1-type $\pi$ realized more than once but no more than $Z$
  times, colour the elements having 1-type $\pi$ using $Z$ different
  colours.  By interpreting the $\log Z $ (rounded up) new unary
  predicates to encode these colours, we obtain the desired expansion
  $\fA'$. This process clearly preserves chromaticity.
\end{proof}
\begin{definition}
  Let $\Sigma$ be a counting signature, let $\fA$ be a structure
  interpreting $\Sigma$, and let $\pi$, $\pi'$ be 1-types over
  $\Sigma$. We say that $\pi$ and $\pi'$ form a {\em noisy pair} in
  $\fA$ if there do not exist distinct $a, a' \in A$ such that ${\rm
    tp}^\fA[a] = \pi$, ${\rm tp}^\fA[a'] = \pi'$, and ${\rm
    tp}^\fA[a,a']$ is silent.
\label{def:fused}
\end{definition}
Thus, if $\pi$ and $\pi'$ form a noisy pair, every element with 1-type
$\pi$ either sends a message to, or receives a message from, every
element with 1-type $\pi'$. Note that Definition~\ref{def:fused} does
not require $\pi$ and $\pi'$ to be distinct.
\begin{lemma}
  Let $\Sigma$ be a counting signature with exactly $m$ counting
  predicates, let $Y$ be a non-negative integer, let $\fA$ be a
  $Y$-branching structure interpreting $\Sigma$, and let $\pi,
  \pi'$ be 1-types over $\Sigma$. If $\pi$ and $\pi'$ form a noisy
  pair in $\fA$, then either there are at most $(mY+1)^2$ elements
  having 1-type $\pi$, or there are at most $(mY+1)^2$ elements having
  1-type $\pi'$.
\label{lma:fusion}
\end{lemma}
\begin{proof}
  Suppose for contradiction that $\pi$ and $\pi'$ form a noisy pair,
  but that there are more than $(mY+1)^2$ elements having 1-type $\pi$
  and more than $(mY+1)^2$ elements having 1-type $\pi'$.  Now let $B$
  be a set of elements having 1-type $\pi$, with $|B| > (mY)^2 + mY$;
  and let $B'$ be a set of elements having 1-type $\pi'$, disjoint
  from $B$, with $|B'| = mY+1$. Such sets can evidently be found even
  if $\pi = \pi'$.  Select any $b \in B$. Since $b$ sends a message to
  at most $mY$ elements in $B'$, there exists $b' \in B$ such that $b$
  sends no message to $b'$. Since $\pi$ and $\pi'$ form a noisy pair,
  we have that, for all $b \in B$, there exists $b' \in B'$ such that
  $b'$ sends a message to $b$.  But the total number of elements $b
  \in A$ such that $b'$ sends a message to $b$ for some $b' \in B'$ is
  bounded by $mY|B'| = mY(mY+1)$, contradicting the supposition that
  $|B| > (mY)^2 + mY$.
\end{proof}

\section{Profiles and approximations}
\label{sec:sparse}
Let $\Sigma$ be a finite counting signature with counting predicates
$f_1, \ldots, f_m$. Enumerate the invertible message-types over
$\Sigma$ (in some arbitrary order) as \linebreak
$\mu_1, \ldots, \mu_{M^*}$, and the non-invertible message-types over
$\Sigma$ (again in some arbitrary order) as $\mu_{M^*+1}, \ldots,
\mu_M$. (Thus, $\mu_1, \ldots, \mu_M$ is an enumeration of all the
message-types over $\Sigma$.)  Denote the set of silent 2-types over
$\Sigma$ by $\Xi$.  We fix the symbols $\Sigma$, $m$, $f_h$ ($1 \leq h
\leq m$), $M^*$, $M$, $\mu_j$ ($1 \leq j \leq M$) and $\Xi$ to have
these meanings throughout this section.

We introduce the concepts of the $\Pi$-{\em profile} and the
$\Pi$-{\em count} of an element in a structure interpreting $\Sigma$,
where $\Pi$ is a set of 1-types.
\begin{notation}
  Let $\fA$ be a structure interpreting $\Sigma$, and let $\pi$ be any
  1-type and $\Pi$ any set of 1-types over $\Sigma$.  When $\fA$ is
  clear from context, denote by $A_\pi$ the set $\{ a \in A| {\rm
  tp}^\fA[a] = \pi \}$, and denote by $A_\Pi$ the set $\bigcup \{
  A_\pi | \pi \in \Pi \}$.  In addition, denote by $\Pi^c$ the set of
  all and only those 1-types over $\Sigma$ not contained in $\Pi$.
\label{notation:A_pi}
\end{notation}
\begin{remark}
  For any $\fA$ interpreting $\Sigma$ and any set of 1-types $\Pi$
  over $\Sigma$, $A_{\Pi^c} = A \setminus A_\Pi$.
\label{remark:PiC}
\end{remark}
For the next definition, recall that $\mu_1, \ldots \mu_M$ are the
message-types (invertible and non-invertible) over $\Sigma$.
\begin{definition}
  Let $\fA$ be a finitely branching structure interpreting $\Sigma$,
  let $a \in A$, and let $\Pi$ be any set of 1-types over $\Sigma$.
  The $\Pi$-{\em profile} of $a$ in $\fA$, denoted ${\rm
    pr}_\Pi^\fA[a]$, is the $M$-element integer vector whose $j$th
  element ($1 \leq j \leq M$) is given by:
\begin{equation*}
| \{ b \in A_\Pi : b \neq a \mbox{ and } {\rm tp}^\fA[a,b] = \mu_j \} |.
\end{equation*}
If $\Pi$ is the set of all 1-types over $\Sigma$, we call ${\rm
  pr}_\Pi^\fA[a]$ simply the {\em profile} of $a$ in $\fA$, and denote
it ${\rm pr}^\fA[a]$.
\label{def:profile}
\end{definition}
Think of the vector ${\rm pr}^\fA[a]$ as a description of the `local
environment' of $a$. The intuition is that $a$'s local environment
tells us, for each message-type $\mu$, {\em how many} elements $a$ is
related to by $\mu$; but it tells us nothing about how those elements
are related to each other, or about the elements of $A$ to which $a$
does not send a message.  More generally, ${\rm pr}_\Pi^\fA[a]$ is a
partial description of that local environment---one that ignores
elements whose 1-type is not in the set $\Pi$.

For the next definition, recall that $f_1, \ldots, f_m$ are the
counting predicates of $\Sigma$.
\begin{definition}
  Let $\fA$ be a finitely branching structure interpreting $\Sigma$,
  let $a \in A$, and let $\Pi$ be any set of 1-types over $\Sigma$.
  The $\Pi$-{\em count} of $a$ in $\fA$, denoted ${\rm
  ct}_\Pi^\fA[a]$, is the $m$-element integer vector whose $h$th
  element ($1 \leq h \leq m$) is given by:
\begin{equation*}
   | \{ b \in A_\Pi : b \neq a  \mbox{ and } \fA \models f_h[a,b] \} |.
\end{equation*}
Furthermore, a set $B \subseteq A$ is called a $\Pi$-{\em group} if
every element of $B$ has the same 1-type and every element of $B$ has
the same $\Pi$-count.  If $\Pi$ is the set of all 1-types over
$\Sigma$, we call ${\rm ct}_\Pi^\fA[a]$ simply the {\em count} of $a$
in $\fA$, and denote it ${\rm ct}^\fA[a]$.
\label{def:Picount}
\end{definition}

Think of the vector ${\rm ct}_\Pi^\fA[a]$ as providing a `statistical
summary' of the vector ${\rm pr}_\Pi^\fA[a]$.  In particular, ${\rm
pr}_\Pi^\fA[a]$ determines ${\rm ct}_\Pi^\fA[a]$, but not
conversely. 
\begin{definition}
  Let $\fA$ be a finitely branching, chromatic structure interpreting
  $\Sigma$, let $\Pi$ be a set of 1-types over $\Sigma$, and let $B$
  be a subset of $A$.  A structure $\fA'$ over the domain $A$ is a
  $(\Pi,B)$-{\em approximation to} $\fA$ if (i) $\fA'$ is chromatic;
  (ii) every 2-type realized in $\fA'$ is also realized in $\fA$; and
  (iii) for all $a \in A$:
\begin{enumerate}
\item ${\rm tp}^{\fA'}[a] = {\rm tp}^\fA[a]$; 
  \label{item:PiBApproximation1}
\item ${\rm pr}^{\fA'}_{\Pi^c}[a] = {\rm
  pr}^\fA_{\Pi^c}[a]$.  
  \label{item:PiBApproximation2}
\item $a \in A \setminus B$ implies ${\rm pr}^{\fA'}[a] = {\rm
  pr}^\fA[a]$; 
  \label{item:PiBApproximation3}
\item $a \in B$ implies
  ${\rm ct}_\Pi^{\fA'}[a] = {\rm ct}_\Pi^\fA[a]$.
  \label{item:PiBApproximation4}
\end{enumerate}
\label{def:PiBApproximation}
\end{definition}
In Condition~\ref{item:PiBApproximation4} of the above definition, the
restriction that $a \in B$ is in fact logically redundant; since if $a
\not \in B$, Condition~\ref{item:PiBApproximation3} certainly entails
${\rm ct}_\Pi^{\fA'}[a] = {\rm ct}_\Pi^\fA[a]$.
\begin{remark}
  If $\fA'$ is a $(\Pi,B)$-approximation to $\fA$, $\Pi \subseteq
  \Pi'$, and $B \subseteq B'$, then $\fA'$ is also a
  $(\Pi',B')$-approximation to $\fA$.  If $\fA'$ is a
  $(\Pi,B)$-approximation to $\fA$, and $\fA''$ is a
  $(\Pi,B)$-approximation to $\fA'$, then $\fA''$ is a
  $(\Pi,B)$-approximation to $\fA$.
\label{remark:PiBapproximation}
\end{remark}

Our strategy now is to show that, in favourable circumstances,
$(\Pi,B)$-approximations can be obtained in which the elements of $B$
exhibit `few' profiles. We first consider the special case where $\Pi$
is a singleton.  For the next definition, recall that $\mu_1, \ldots,
\mu_{M^*}$ are the invertible message-types over $\Sigma$; thus, the
first $M^*$ coordinates of any $\pi$-profile ${\rm
  pr}_{\{\pi\}}^\fA[a]$ tell us, for each invertible message-type
$\mu$, how many elements with 1-type $\pi$ $a$ sends a message of type
$\mu$ to.
\begin{definition}
  Let $\pi$ be a 1-type over $\Sigma$, let $\fA$ be a finitely
  branching structure interpreting $\Sigma$, and let $B$ be a subset
  of $A$.  We say that $B$ is a $\pi$-{\em patch} if $B$ is a $\{ \pi
  \}$-group and, for all $a, b \in B$, the vectors ${\rm
  pr}_{\{\pi\}}^\fA[a]$ and ${\rm pr}_{\{\pi\}}^\fA[b]$ agree in each
  of their first $M^*$ coordinates.
\label{def:pipatch}
\end{definition}
\begin{lemma}
  Let $\fA$ be a chromatic, $Y$-branching structure interpreting
  $\Sigma$, let $\pi$ be a 1-type over $\Sigma$, and let $B \subseteq
  A$ be a $\pi$-patch. If $\fA$ is $(mY+1)^2$-differentiated, then
  there exists a structure $\fA'$ such that $\fA'$ is a $(\{\pi \},
  B)$-approximation to $\fA$ in which the elements of $B$ all have the
  same $\{ \pi \}$-profile.
\label{lma:bBpi}
\end{lemma}
\begin{proof}
  Let $\pi^*$ be the 1-type such that $B \subseteq A_{\pi^*}$, and
  suppose that $\fA$ is $(mY+1)^2$-differentiated.  If $|B| \leq 1$,
  $\fA' = \fA$ obviously satisfies the conditions of the Lemma.  And
  if $|A_\pi| \leq 1$, any two elements with the same $\{\pi\}$-count
  have the same $\{\pi\}$-profile, so that $\fA' = \fA$ again
  satisfies the conditions of the Lemma.  So we may suppose that $B$
  and $A_\pi$ both contain more than one element, whence $|A_{\pi^*}|
  > (mY+1)^2$ and $|A_\pi| > (mY+1)^2$.  By
  Lemma~\ref{lma:fusion}, let $\tau$ be a silent 2-type such that, for
  some $a \in A_{\pi^*}$ and some $a' \in A_\pi$, ${\rm tp}^\fA[a, a']
  = \tau$.

\vspace{0.25cm}

\noindent
  For $a \in B$, let
\begin{equation*}
  A_a = \{a' \in A_\pi \mid \mbox{ $a \neq a'$ and ${\rm tp}^\fA[a,a']$ 
                                   is a non-invertible message-type} \};
\end{equation*}
and for $a \not \in B$, let $A_a = \emptyset$.  Notice, incidentally,
that $a' \in A_a$ implies $a \not \in A_{a'}$.  Choose $b \in B$ for
which $|A_b|$ is smallest, and fix $b$. Enumerate $A_b$ (finite or
infinite) as $b_1, b_2, \ldots$~. For any $a \in B$ not equal to $b$,
let $\hat{A}_a$ be a subset of $A_a$ having the same number of
elements as $A_b$, and enumerate $\hat{A}_a$ as $a_1, a_2, \ldots$~.

\vspace{0.25cm}

\noindent
We now define the structure $\fA'$ by assigning 2-types as follows.
For all $a \in B$ such that $ a \neq b$, set
\begin{equation}
{\rm tp}^{\fA'}[a,a_i] = {\rm tp}^\fA[b,b_i],
\label{eq:bBpitransform1}
\end{equation}
where $i$ ranges over the enumeration of $\hat{A}_a$, and set
\begin{equation}
{\rm tp}^{\fA'}[a,a'] = \tau,
\label{eq:bBpitransform2}
\end{equation}
where $a'$ is any element of $A_a \setminus \hat{A}_a$.  In
addition, for all distinct $a$, $a'$ such that $a' \not \in A_a$ and
$a \not \in A_{a'}$, set
\begin{equation}
{\rm tp}^{\fA'}[a,a'] = {\rm tp}^\fA[a,a'].
\label{eq:bBpitransform3}
\end{equation}
Since $a' \in A_a$ implies $a \not \in A_{a'}$, none of these
assignments overwrites any other. And since $B \subseteq A_{\pi^*}$,
the 1-type assignments implicit
in~\eqref{eq:bBpitransform1}--\eqref{eq:bBpitransform3} never clash:
indeed, we have ${\rm tp}^{\fA'}[a] = {\rm tp}^\fA[a]$ for all $a \in
A$. Furthermore, the transformation from $\fA$ to $\fA'$ does not
affect invertible message-types. That is: for distinct $a$, $a'$,
${\rm tp}^\fA[a,a']$ is an invertible message-type if and only if
${\rm tp}^{\fA'}[a,a']$ is an invertible message-type; and moreover,
if ${\rm tp}^\fA[a,a']$ is an invertible message-type, then ${\rm
  tp}^{\fA'}[a,a']= {\rm tp}^\fA[a,a']$. 

\vspace{0.25cm}

\noindent
We now verify that $\fA'$ is a $(\{ \pi \},B)$-approximation to $\fA$.
From the remarks of the previous paragraph and the fact that $\fA$ is
chromatic, we have that $\fA'$ is also chromatic. In addition, it is
immediate from~\eqref{eq:bBpitransform1}--\eqref{eq:bBpitransform3}
that every 2-type realized in $\fA'$ is also realized in $\fA$.  Now
let $a$ be any element of $B$.  Since $B$ is a $\pi$-patch, the
vectors ${\rm pr}_{\{\pi\}}^{\fA}[a]$ and ${\rm
  pr}_{\{\pi\}}^{\fA}[b]$ by definition agree in their first $M^*$
coordinates (corresponding to the invertible message-types). Hence,
since we have just shown that the transformation from $\fA$ to $\fA'$
does not affect invertible message-types, the vectors ${\rm
  pr}_{\{\pi\}}^{\fA'}[a]$ and ${\rm pr}_{\{\pi\}}^{\fA}[b]$ also
agree in their first $M^*$ coordinates. Furthermore, the
assignments~\eqref{eq:bBpitransform1} and~\eqref{eq:bBpitransform2}
guarantee that ${\rm pr}_{\{\pi\}}^{\fA'}[a]$ and ${\rm
  pr}_{\{\pi\}}^{\fA}[b]$ also agree in the remaining coordinates
$M^*+1, \ldots, M$ (corresponding to the non-invertible
message-types).  Hence we have:
\begin{equation}
{\rm pr}_{\{\pi\}}^{\fA'}[a] = {\rm pr}_{\{\pi\}}^\fA[b] \qquad
\text{for all $a \in B$}.
\label{eq:preserved}
\end{equation}
It is now a simple matter to check the numbered conditions in
Definition~\ref{def:PiBApproximation} (iii). Let $a \in A$.
\begin{enumerate}
\item We have already established that ${\rm tp}^{\fA'}[a] = {\rm
tp}^\fA[a]$.
\item 
Let $a'$ be any element of $A_{\{\pi\}^c}$ with $a \neq a'$.  Then
certainly $a' \not \in A_a \subseteq A_\pi$, so that ${\rm
tp}^{\fA'}[a,a']$ can be different from ${\rm tp}^\fA[a,a']$ only if
$a \in A_{a'}$. But if $a \in A_{a'}$, then neither ${\rm
tp}^\fA[a,a']$ nor ${\rm tp}^{\fA'}[a,a']$ can be a message-type.
Hence, ${\rm pr}_{\{\pi\}^c}^{\fA'}[a] = {\rm
pr}_{\{\pi\}^c}^\fA[a]$.
\item Suppose $a \in A \setminus B$, and let $a'$ be any element of
  $A$ with $a \neq a'$. The argument now proceeds much as for the
  previous condition: certainly, $a' \not \in A_a = \emptyset$, so
  ${\rm tp}^{\fA'}[a,a']$ can be different from ${\rm tp}^\fA[a,a']$
  only if $a \in A_{a'}$.  But if $a \in A_{a'}$, then neither ${\rm
    tp}^\fA[a,a']$ nor ${\rm tp}^{\fA'}[a,a']$ can be a message-type.
  Hence ${\rm pr}^{\fA'}[a] = {\rm pr}^\fA[a]$.
\item If $a \in B$, Equation~\eqref{eq:preserved} yields ${\rm
ct}_{\{\pi\}}^{\fA'}[a] = {\rm ct}_{\{\pi\}}^\fA[b]$. And since $B$ is
a $\{\pi\}$-group, ${\rm ct}_{\{\pi\}}^{\fA}[b] = {\rm
ct}_{\{\pi\}}^\fA[a]$. It follows that ${\rm ct}_{\{\pi\}}^{\fA'}[a] =
{\rm ct}_{\{\pi\}}^\fA[a]$.
\end{enumerate}
Finally, it is immediate from Equation~\eqref{eq:preserved} that all
elements of $B$ have the same $\{\pi\}$-profile in $\fA'$.
\end{proof}
Thus, Lemma~\ref{lma:bBpi} assures us that, as long as we are content
to work with $(\{ \pi \},B)$-approximations to highly differentiated
structures, we may unify the $\{ \pi \}$-profiles of the elements in
any $\pi$-patch $B$. 

We next deal with the case where $\Pi$ is not a singleton. Here, the
possibilities for reducing the number of $\Pi$-profiles realized in a
$\Pi$-group are more limited. 
\begin{lemma}
  Let $\fA$ be a finitely branching structure interpreting $\Sigma$,
  let $\Pi$ be a set of 1-types over $\Sigma$, let $B \subseteq A$ be
  a $\Pi$-group, and let $\omega$ be a permutation of $B$. Then there
  exists a structure $\fA'$ interpreting $\Sigma$ over the domain $A$
  such that $\fA'$ is a $(\Pi , B)$-approximation to $\fA$, and
  for all $b \in B$, ${\rm pr}_\Pi^{\fA'}[\omega(b)] = {\rm
  pr}_\Pi^\fA[b]$.
\label{lma:permute}
\end{lemma}
\begin{proof}
  First, extend $\omega$ to the whole of $A$ by setting $\omega(a) =
  a$ for $a \in A \setminus B$. Next, for all $b \in A$, define:
\begin{equation*}
\omega_{b\Pi}(a) =
\begin{cases}
  \omega(a) \text{ if $b \in A_\Pi$}\\
  a \text{ otherwise};
\end{cases}
\end{equation*}
Thus, $\omega_{b\Pi}$ is a permutation of $A$ (which may be the
identity). Since $(\omega_{b\Pi})^{-1}$ and $(\omega^{-1})_{b\Pi}$ are
the same permutation, we may unambiguously write $\omega_{b\Pi}^{-1}$.
Clearly, $\omega$ fixes $B$ setwise and $A \setminus B$ pointwise; so,
therefore, does $\omega^{-1}_{b\Pi}$. Moreover, since $B$ is a
$\Pi$-group, either $B \subseteq A_\Pi$ or $B \subseteq A_{\Pi^c} = A
\setminus A_\Pi$. Hence, $\omega$ fixes both $A_\Pi$ and $A_{\Pi^c}$
setwise, and so therefore does $\omega_{b\Pi}^{-1}$.

\vspace{0.25cm}

\noindent
Define the structure $\fA'$ over domain $A$ by setting, for all
distinct $a, a' \in A$:
\begin{equation}
{\rm tp}^{\fA'}[a,a'] = 
      {\rm tp}^\fA[\omega^{-1}_{a'\Pi}(a), \omega^{-1}_{a\Pi}(a')].
\label{eq:switch}
\end{equation}
To show that $\fA'$ is well-defined, we must show first that the
elements $\omega^{-1}_{a'\Pi}(a)$ and $\omega^{-1}_{a\Pi}(a')$ in each
instance of~\eqref{eq:switch} are distinct, and second, that the
1-type assignments implicit in the different instances
of~\eqref{eq:switch} do not clash.  Suppose, then that $a \neq a'$; we
prove that $\omega^{-1}_{a'\Pi}(a) \neq \omega^{-1}_{a\Pi}(a')$. Since
the permutations $\omega^{-1}_{a\Pi}$ and $\omega^{-1}_{a'\Pi}$ fix
$B$ setwise and $A \setminus B$ pointwise, we may assume that $a, a'
\in B$. If $B \subseteq A_\pi$, then $\omega^{-1}_{a'\Pi}(a) =
\omega^{-1}(a)$ and $\omega^{-1}_{a\Pi}(a')= \omega^{-1}(a')$; if, on
the other hand, $B \subseteq A \setminus A_\pi$, then
$\omega^{-1}_{a'\Pi}(a) = a$ and $\omega^{-1}_{a\Pi}(a') = a'$. Either
way, $\omega^{-1}_{a'\Pi}(a) \neq \omega^{-1}_{a\Pi}(a')$. Next, we
prove that the 1-type assignments in~\eqref{eq:switch} never clash.
Since all elements of $B$ have the same 1-type, and since $\omega$ is
the identity outside $B$, we have, for all $a, a'$, ${\rm
tp}^{\fA}[\omega^{-1}_{a'\Pi}(a)]= {\rm tp}^{\fA}[a]$; thus, ${\rm
tp}^{\fA}[\omega^{-1}_{a'\Pi}(a)]$ does not depend on $a'$.  Hence,
the 1-type assignments implicit in~\eqref{eq:switch} cannot clash, and
$\fA'$ is indeed well-defined. In fact, this argument establishes that
${\rm tp}^{\fA'}[a] = {\rm tp}^{\fA}[a]$ for all $a \in A$.

\vspace{0.25cm}

\noindent
We first check the numbered
conditions of Definition~\ref{def:PiBApproximation} (iii)
in turn. Let $a
\in A$.
\begin{enumerate}
\item We have just established that ${\rm tp}^{\fA'}[a] = {\rm
tp}^{\fA}[a]$.
\item 
For all $b \in A_{\Pi^c}$, $\omega^{-1}_{b\Pi}(a) = a$; in
particular, if $a \in A_{\Pi^c}$, then $\omega^{-1}_{a\Pi}(a) = a$.
Therefore, $\omega^{-1}_{a\Pi}$ is in fact a permutation of $A_{\Pi^c}
\setminus \{ a \}$, and moreover, for all $b \in A_{\Pi^c} \setminus
\{ a \}$, ${\rm tp}^{\fA'}[a,b] = {\rm
tp}^\fA[a,\omega^{-1}_{a\Pi}(b)]$. Thus, the list of 2-types ${\rm
tp}^{\fA'}[a,b]$ obtained as $b$ ranges over $A_{\Pi^c} \setminus \{ a
\}$ is (in some order) the list of 2-types ${\rm tp}^\fA[a,b']$
obtained as $b'$ ranges over $A_{\Pi^c} \setminus \{ a \}$.  It
follows that ${\rm pr}^{\fA'}_{\Pi^c}[a] = {\rm pr}^\fA_{\Pi^c}[a]$.
\item
If $a \in A \setminus B$, then, for all $b \in A$,
$\omega_{b\Pi}^{-1}(a) = a$; in particular,
$\omega^{-1}_{a\Pi}(a) = a$.  Therefore,
$\omega^{-1}_{a\Pi}$ is a permutation of $A \setminus \{ a \}$, and
moreover, for all $b \in A \setminus \{ a \}$, ${\rm tp}^{\fA'}[a,b] =
{\rm tp}^\fA[a,\omega^{-1}_{a\Pi}(b)]$. Thus, the list of 2-types
${\rm tp}^{\fA'}[a,b]$ obtained as $b$ ranges over $A \setminus \{ a
\}$ is (in some order) the list of 2-types ${\rm tp}^\fA[a,b']$
obtained as $b'$ ranges over $A \setminus \{ a \}$.  It follows that
${\rm pr}^{\fA'}[a] = {\rm pr}^\fA[a]$.
\item
For all $b \in A_\Pi$, $\omega^{-1}_{b\Pi}(a) = \omega^{-1}(a)$; in
particular, if $a \in A_\Pi$, then $\omega^{-1}_{a\Pi}(a) =
\omega^{-1}(a)$.  Therefore, $\omega^{-1}_{a\Pi}$ is a bijection from
the set $A_\Pi \setminus \{ a \}$ to the set $A_\Pi \setminus \{
\omega^{-1}(a) \}$, and moreover, for all $b \in A_\Pi \setminus \{ a
\}$, ${\rm tp}^{\fA'}[a,b] = {\rm
tp}^\fA[\omega^{-1}(a),\omega^{-1}_{a\Pi}(b)]$.  Thus, the list of
2-types ${\rm tp}^{\fA'}[a,b]$ obtained as $b$ ranges over $A_\Pi
\setminus \{ a \}$ is (in some order) the list of 2-types ${\rm
tp}^\fA[\omega^{-1}(a),b']$ obtained as $b'$ ranges over $A_\Pi
\setminus \{ \omega^{-1}(a) \}$.  It follows that
\begin{equation}
{\rm pr}_\Pi^{\fA'}[a] = {\rm pr}_\Pi^\fA[\omega^{-1}(a)].
\label{eq:permute}
\end{equation}
Certainly, then, we have ${\rm ct}_\Pi^{\fA'}[a] = {\rm
ct}_\Pi^\fA[\omega^{-1}(a)]$.  But since $B$ is a $\Pi$-group in
$\fA$, ${\rm ct}_\Pi^\fA[\omega^{-1}(a)]= {\rm ct}_\Pi^\fA[a]$. Hence
${\rm ct}_\Pi^{\fA'}[a] = {\rm ct}_\Pi^\fA[a]$.
\end{enumerate}
We have thus established that, for all $a \in A$, ${\rm
pr}_\Pi^{\fA'}[a] = {\rm pr}_\Pi^\fA[\omega^{-1}(a)]$ and ${\rm
pr}_{\Pi^c}^{\fA'}[a] = {\rm pr}_{\Pi^c}^\fA[a]$.  Since $\fA$ is
chromatic, it follows easily that $\fA'$ is chromatic and that all
2-types realized in $\fA'$ are realized in $\fA$. Hence, $\fA'$ is a
$(\Pi,B)$-approximation to $\fA$.  Finally, it follows by putting
$a = \omega(b)$ in
Equation~\eqref{eq:permute} that, for all $b \in B$, ${\rm
pr}_\Pi^{\fA'}[\omega(b)] = {\rm pr}_\Pi^\fA[b]$.
\end{proof}
Let $\fA$, $\Pi$ and $B$ satisfy the conditions of
Lemma~\ref{lma:permute}. That lemma then assures us that, as long as
we are content to work with $(\Pi,B)$-approximations, we may permute
the $\Pi$-profiles of the elements in any $\Pi$-group $B$ at will!  To
see the power of this idea, let $\fA$ be a structure, let $\Pi'$ and
$\Pi''$ be disjoint sets of 1-types, and let $B$ be a subset of $A$
whose elements realize $J'$ different $\Pi'$-profiles and $J''$
different $\Pi''$-profiles.  If $\Pi = \Pi' \cup \Pi''$, how many
$\Pi$-profiles are realized by the elements of $B$?  Answer: in
general $J'J''$; however, the following lemma shows that, under
certain circumstances, we can find a $(\Pi,B)$-approximation to $\fA$
for which the answer is smaller.
\begin{lemma}
$\fA$ be a finitely branching structure interpreting $\Sigma$, let
$\Pi'$, $\Pi''$ be disjoint, nonempty sets of 1-types, and let $\Pi =
\Pi' \cup \Pi''$.  Suppose $B \subseteq A$ is both a $\Pi'$-group and
a $\Pi''$-group, and hence also a $\Pi$-group. Let the number of
different $\Pi'$-profiles realized in $\fA$ by the elements of $B$ be
$J'$; and let the number of different $\Pi''$-profiles realized in
$\fA$ by the elements of $B$ be $J''$.  Then there exists a
$(\Pi,B)$-approximation $\fA''$ to $\fA$ in which at most $J' + J''$
different $\Pi$-profiles are realized by the elements of $B$.
\label{lma:profilegrouping}
\end{lemma}
\begin{proof}
  For perspicuity, we give the proof where $B$ is finite; the
  modifications required for the general case are easy. Enumerate $B$
  as $b_1, \ldots b_I$. Let the various $\Pi'$-profiles realized by at
  least one element of $B$ be $\bar{v}'_1, \ldots, \bar{v}'_{J'}$; and
  let the various $\Pi''$-profiles realized by at least one element of
  $B$ be $\bar{v}''_1, \ldots, \bar{v}''_{J''}$.  Since $B$ is a
  $\Pi'$-group, Lemma~\ref{lma:permute} guarantees that we can obtain
  a $(\Pi',B)$-approximation to $\fA$ in which the $\Pi'$-profiles of
  $B$ are permuted at will. So let $\fA'$ be a
  $(\Pi',B)$-approximation to $\fA$ in which the $\Pi'$-profiles of
  the $b_1, \ldots b_I$ fall into consecutive blocks in the sense
  depicted in the middle column in Fig.~\ref{fig:profilegrouping}.
  More precisely, we have integers $0 = I_1 < I_2 < \cdots < I_{J'+1}
  = I$ such that, for all $j$ ($1 \leq j \leq J'$), ${\rm
  pr}^{\fA'}[b_i] = \bar{v}'_j$ for $i$ in the range $[I_j +1,
  I_{j+1}]$.  Since $B$ is a also $\Pi''$-group, we can likewise
  obtain a structure $\fA''$ such that $\fA''$ is a
  $(\Pi'',B)$-approximation to $\fA'$ in which the elements of $B$
  have $\Pi''$-profiles likewise arranged in consecutive blocks. Since
  $\fA''$ is a $(\Pi'',B)$-approximation to $\fA'$ and the sets $\Pi'$
  and $\Pi''$ are disjoint, the $\Pi'$-profiles of the elements of $B$
  will be unaffected by the transformation from $\fA'$ to $\fA''$: a
  typical alignment of $\Pi'$-profiles and $\Pi''$-profiles in $\fA''$
  is shown in Fig.~\ref{fig:profilegrouping}.
\begin{figure}
\begin{center}
\begin{tabular}{|c|c|c|}
\hline
\ Element & $\Pi'$-profile in $\fA'$ & $\Pi''$-profile in $\fA''$\\
\ of $B$  & (and also in $\fA''$)    & \ \\
\hline
$b_1$ & \ & \ \\
\ & $\bar{v}'_1$ & $\bar{v}''_1$ \\
\cline{3-3}
\ & \ & \ \\
\cline{2-2}
\ & \ & $\bar{v}''_2$ \\
\cline{3-3}
\ & $\bar{v}'_2$ & \ \\
\cline{2-2}
\ & \ & \ \\
\ & \vdots & \vdots \\
\ & \ & \ \\
\cline{3-3}
\ & \ & \ \\
\cline{2-2}
\ & \ & $\bar{v}''_{J''}$ \\
$b_I$ & $\bar{v}'_{J'}$ & \ \\
\hline
\end{tabular}
\end{center}
\caption{Arrangement of $\Pi'$-profiles and $\Pi''$-profiles in $B$.}
\label{fig:profilegrouping}
\end{figure}
By inspection, at most $J'$+$J''$ $\Pi$-profiles are realized in
$\fA''$ by the elements of $B$.  From
Remark~\ref{remark:PiBapproximation}, $\fA''$ is a
$(\Pi,B)$-approximation to $\fA$, because $\fA'$ is a
$(\Pi',B)$-approximation to $\fA$ and $\fA''$ is a
$(\Pi'',B)$-approximation to $\fA'$.
\end{proof}

This leads us to the main result of this section. 
\begin{lemma}
  Let $\fA$ be a $Y$-branching, $(mY+1)^2$-differentiated, chromatic
  structure over $\Sigma$, let $l \geq 0$, let $\Pi$ be a non-empty
  set of 1-types such that $| \Pi | \leq 2^l$, and let $B \subseteq A$
  be a $\Pi$-group.  Then there is a structure $\fA'$ such that $\fA'$
  is a $(\Pi,B)$-approximation to $\fA$ and the number of different
  $\Pi$-profiles realized in $\fA'$ by the elements of $B$ is at most
  $2^l(M^*+1)(Y+1)^{lm}$.
\label{lma:fewprofiles}
\end{lemma}
\begin{proof}
  By induction on $l$.  To aid readability, let $K_l$ stand for
  $2^l(M^*+1)(Y+1)^{lm}$. If $l = 0$, let $\Pi = \{ \pi \}$.  Decompose
  $B$ into maximal $\pi$-patches $B_1, \ldots B_H$. Since $\fA$ is
  chromatic, the first $M^*$ coordinates of ${\rm
  pr}_{\{\pi\}}^\fA[a]$ are either all zero, or else are all zero
  except for a single occurrence of unity. Therefore, $H \leq M^*+1$.
  Now let $\fA_0 = \fA$, and for all $h$ ($1 \leq h \leq H$), apply
  Lemma~\ref{lma:bBpi} to obtain a structure $\fA_h$ such that
  $\fA_h$ is a $(\{\pi \}, B_h)$-approximation to $\fA_{h-1}$ in which
  the elements of $B_h$ all have the same $\{ \pi \}$-profile.  By
  Remark~\ref{remark:PiBapproximation}, $\fA_H$ is a
  $(\{\pi\},B)$-approximation to $\fA$. And because the $B_h$ are
  pairwise disjoint, $1 \leq h < h' \leq H$ implies ${\rm
  pr}^{\fA_{h'}}[a] = {\rm pr}^{\fA_{h}}[a]$ for all $a \in B_h$.
  Hence, the total number of $\{ \pi \}$-profiles realized by elements
  of $B$ in $\fA_H$ is at most $H \leq M^*+1 =
  K_0$. Thus, setting $\fA' = \fA_H$ establishes the
  case $l = 0$.

\vspace{0.25cm}

\noindent
Now suppose $l > 0$. We may assume $\Pi$ is not a singleton, since
otherwise, we can employ the argument of the case $l=0$; so let $\Pi$
be partitioned into non-empty sets $\Pi'$ and $\Pi''$ each of
cardinality at most $2^{l-1}$. Also, partition $B$ into maximal
$\Pi'$-groups $B_1$, \ldots, $B_H$ (say). Since $B$ is a $\Pi$-group
and $\Pi' \cap \Pi'' = \emptyset$, the $B_1$, \ldots, $B_H$ will also
be $\Pi''$-groups.  Moreover, since $\fA$ is $Y$-branching, the
$\Pi$-count of any element in $\fA$ is one of at most $(Y+1)^m$
different vectors; and since $B$ is a $\Pi$-group, all elements of $B$
must have the same 1-type, whence $H \leq (Y+1)^m$.  Again, let $\fA_0
= \fA$, and consider the set $B_1$.  By inductive hypothesis, let
$\fA_1'$ be a $(\Pi',B_1)$-approximation to $\fA_0$ in which at most
$K_{l-1}$ $\Pi'$-profiles are realized by the elements of $B_1$.
Again, by inductive hypothesis, let $\fA_1''$ be a
$(\Pi'',B_1)$-approximation to $\fA_1'$ in which at most $K_{l-1}$
$\Pi''$-profiles are realized by the elements of $B_1$. Thus, in the
structure $\fA_1''$, $B_1$ is a $\Pi'$-group realizing at most
$K_{l-1}$ different $\Pi'$-profiles, and also a $\Pi''$-group
realizing at most $K_{l-1}$ different $\Pi''$-profiles. By
Lemma~\ref{lma:profilegrouping}, let $\fA_1$ be a
$(\Pi,B_1)$-approximation to $\fA''_1$ in which the elements of $B_1$
realize at most $2K_{l-1}$ different $\Pi$-profiles.  By
Remark~\ref{remark:PiBapproximation}, $\fA_1$ is a
$(\Pi,B_1)$-approximation to $\fA_0$.  Treating the sets $B_2$,
\ldots, $B_H$ in the same way, we obtain structures $\fA_h$ ($1 \leq h
\leq H$) such that, for each $h$ in this range, $\fA_h$ is a
$(\Pi,B)$-approximation to $\fA_{h-1}$ in which at most $2K_{l-1}$
different $\Pi$-profiles are realized by the elements of $B_h$.  By
Remark~\ref{remark:PiBapproximation}, $\fA_H$ is a
$(\Pi,B)$-approximation to $\fA$. And because the $B_h$ are pairwise
disjoint, $1 \leq h < h' \leq H$ implies that ${\rm pr}^{\fA_{h'}}[a]
= {\rm pr}^{\fA_{h}}[a]$ for all $a \in B_h$.  Hence, the total number
of $\Pi$-profiles realized by elements of $B$ in $\fA_H$ is at most
\begin{align*}
2HK_{l-1}  & 
            \leq 2(Y+1)^mK_{l-1}\\
 & = K_l.
\end{align*}
Thus, setting $\fA' = \fA_H$ completes the induction.
\end{proof}

\section{Deciding finite satisfiability}
\label{sec:finsat}
Let $\Sigma$ be a signature of unary and binary predicates, let
$\alpha(x)$ be a quantifier-free, equality-free formula over $\Sigma$
with $x$ as its only variable, let $\beta(x,y)$ be a quantifier-free,
equality-free formula over $\Sigma$ with $x$ and $y$ as its only
variables, let $m$ be a positive integer, let $f_1, \ldots, f_m$ be
binary predicates, let $C_1, \ldots, C_m$ be positive integers, let
$\phi^*$ be the formula
\begin{equation}
\phi^* := 
\forall x \alpha \wedge 
 \forall x \forall y (\beta \vee x \approx y) \wedge 
   \bigwedge_{1 \leq h \leq m} 
      \forall x \exists_{=C_h} y (f_h(x,y) \wedge x \not \approx y),
\label{eq:normalform}
\end{equation}
and let $C = \max_{1 \leq h \leq m} C_h$.  Make $\Sigma$ into a
counting signature by declaring the counting predicates of $\Sigma$ to
be exactly $\{f_1, \ldots, f_m\}$. Let the number of symbols (unary
and binary predicates) in $\Sigma$ be $s$, so that the total number of
1-types over $\Sigma$ is $L = 2^s$.  As in Section~\ref{sec:sparse},
enumerate the invertible message-types over $\Sigma$ (in some
arbitrary order) as $\mu_1, \ldots, \mu_{M^*}$, enumerate the
non-invertible message-types over $\Sigma$ (again in some arbitrary
order) as $\mu_{M^*+1}, \ldots, \mu_M$, and denote the set of silent
2-types over $\Sigma$ by $\Xi$.  We fix the symbols $\alpha$, $\beta$,
$m$, $f_h$ ($1 \leq h \leq m$), $C_h$ ($1 \leq h \leq m$), $C$,
$\phi^*$, $\Sigma$, $s$, $L$, $M^*$, $M$, $\mu_j$ ($1 \leq j \leq M$)
and $\Xi$ to have these meanings throughout this section.

The motivation for introducing $(\Pi,B)$-approximations in
Section~\ref{sec:sparse} is that they are good enough for checking the
(finite) satisfiability of $\phi^*$.
\begin{lemma}
  Let $\fA$ be a finitely branching, chromatic structure interpreting
  $\Sigma$, let $\Pi$ be a set of 1-types over $\Sigma$, let $B$ be a
  subset of $A$, and let $\fA'$ be a $(\Pi,B)$-approximation to
  $\fA$. If $\fA \models \phi^*$, then $\fA' \models \phi^*$.
\label{lma:preservation}
\end{lemma}
\begin{proof}
  By Remark~\ref{remark:PiBapproximation}, we may assume without loss
  of generality that $\Pi$ is the set of all 1-types and $B = A$.
  Since ${\rm tp}^{\fA'}[a] = {\rm tp}^\fA[a]$ for every $a \in A$,
  $\fA' \models \forall x \alpha$.  Since every 2-type realized in
  $\fA'$ is also realized in $\fA$, $\fA' \models \forall x
  \forall y (\beta \vee x \approx y)$. And since ${\rm ct}^{\fA'}[a] =
  {\rm ct}^\fA[a]$ for every $a \in A$, $\fA' \models \bigwedge_{1
  \leq h \leq m} \forall x \exists_{=C_h} y (f_h(x,y) \wedge x \not
  \approx y)$.
\end{proof}

The next definition relies on conventions established in
Notation~\ref{notation:types}.
\begin{definition}
  A {\em star-type} ({\em over} $\Sigma$) is a pair $\sigma = \langle
  \pi, \bar{v} \rangle$, where $\pi$ is a 1-type over $\Sigma$ and
  $\bar{v} = (v_1, \ldots v_M)$ is a vector over $\bbN$ satisfying the
  condition that, for all $j$ ($1 \leq j \leq M$), $v_j >0$ implies
  ${\rm tp}_1(\mu_j) = \pi$. We say that $\sigma$ is {\em chromatic}
  if the set of integers
\[
\{ v_j \mid 1 \leq j \leq M^* \mbox{ and } {\rm tp}_2(\mu_j) = \pi' \}
\]
sums to either 0 or 1 for every 1-type $\pi'$, and sums to 0 in the
case $\pi' = \pi$.  If $\fA$ is a finitely branching structure
interpreting $\Sigma$ and $a \in A$, then $\langle {\rm tp}^\fA[a],
{\rm pr}^\fA[a] \rangle$ is evidently a star-type, which we denote by
${\rm st}^\fA[a]$. We say that the star-type $\sigma$ is {\em
  realized} in $\fA$ if $\sigma = {\rm st}^\fA[a]$ for some $a \in A$.
\label{def:startype}
\end{definition}
We note in passing that, if $\langle \pi, \bar{v} \rangle$ is a
star-type, and $\bar{v}$ is not the zero-vector, then $\pi$ is
actually determined by $\bar{v}$.
\begin{notation}
  If $\sigma = \langle \pi, \bar{v} \rangle$ is a star-type with
  $\bar{v} = (v_1, \ldots v_M)$, we denote $\pi$ by ${\rm tp}(\sigma)$
  and $v_j$ by $\sigma[j]$ for all $j$ ($1 \leq j \leq M$).
\label{notation:startype}
\end{notation}
\begin{remark}
  If $\fA$ is a finitely branching structure interpreting $\Sigma$ and
  $a \in A$, then ${\rm tp}({\rm st}^\fA[a]) = {\rm tp}^\fA[a]$.
  Moreover, $\fA$ is chromatic if and only if every star-type realized
  in $\fA$ is chromatic.
\label{remark:profilevector}
\end{remark}
\begin{remark}
Let $\sigma$ be a chromatic star-type and let $j$ and $j'$ be
integers between 1 and $M^*$ (so that $\mu_j$ and $\mu_{j'}$ are
invertible message-types). If $\mu_j^{-1} = \mu_{j'}$, then either
$\sigma[j] = 0$ or $\sigma[j'] = 0$. In particular, if $\mu_j^{-1} =
\mu_{j}$, then $\sigma[j]= 0$.
\label{remark:chromaticprofilevector}
\end{remark}
\begin{definition}
  Let $\fA$ be a finitely branching structure interpreting $\Sigma$,
  and let $X$ be a positive integer. We say that $\fA$ is $X$-{\em sparse}
  if $\fA$ realizes no more than $X$ different star-types---that is,
  if $| \{ {\rm st}^\fA[a] : a \in A \} | \leq X$.
\label{def:sparse}
\end{definition}
For the next lemma, recall that $s$ is the number of symbols in
$\Sigma$, $m$ is the number of counting predicates in $\Sigma$, and $C
= \max_{1 \leq h \leq m} C_h$.
\begin{lemma}
  Let $Z \geq (mC+1)^2$, and let $X = 4^s(16^s + 1)(C+1)^{sm}$.  If
  there is a chromatic, $Z$-differentiated model of $\phi^*$
  interpreting $\Sigma$, then there is a chromatic,
  $Z$-differentiated, $X$-sparse model of $\phi^*$ interpreting
  $\Sigma$ over the same domain.
\label{lma:sparse}
\end{lemma}
\begin{proof}
  Suppose $\fA$ is a chromatic, $Z$-differentiated model of $\phi^*$
  interpreting $\Sigma$. Since $\fA \models \phi^*$, $\fA$ is
  $C$-branching.  Note that $M^* \leq 2^{4s}$.  Let $\Pi$ be the set
  of all 1-types over $\Sigma$; thus, $|\Pi| = 2^s$.  Let $A_1, \ldots
  A_H$ be a list of the nonempty sets $A_\pi$, where $\pi \in \Pi$;
  thus, $H \leq 2^s$.  The sets $A_h$ together partition $A$;
  moreover, since $\fA \models \phi^*$, each $A_h$ ($1 \leq h \leq H$)
  is a $\Pi$-group.  Letting $\fA_0 = \fA$, by
  Lemma~\ref{lma:fewprofiles}, we can obtain $\fA_1, \ldots \fA_H$
  such that, for all $h$ ($1 \leq h \leq H$), $\fA_h$ is a
  $(\Pi,A_h)$-approximation to $\fA_{h-1}$ in which at most $2^s(16^s
  + 1)(C+1)^{sm}$ different profiles are realized by the elements of
  $A_h$.  Using by now familiar reasoning, $\fA_H$ is therefore a
  $(\Pi,A)$-approximation to $\fA$ realizing at most $H.[2^s(16^s +
  1)(C+1)^{sm}] \leq 4^s(16^s + 1)(C+1)^{sm} = X$ different
  star-types.  By Lemma~\ref{lma:preservation}, $\fA_H \models
  \phi^*$. By Definition~\ref{def:PiBApproximation}, $\fA_H$ is
  chromatic; and since ${\rm tp}^{\fA_H}[a] = {\rm tp}^\fA[a]$ for all
  $a \in A$, $\fA_H$ is also $Z$-differentiated. Thus $\fA' = \fA_H$
  is the required structure.
\end{proof}
\begin{notation}
  We write $\cI$ to denote the set of (unordered) pairs of (not
  necessarily distinct) integers between $1$ and $L$.  Formally: $\cI
  = \{ \{i,i'\} \mid 1 \leq i \leq i' \leq L\}$.
\label{notation:cI}
\end{notation}
For the next definition, recall that $\Xi$ is the set of silent
2-types over $\Sigma$.
\begin{definition}
  A {\em frame} over $\Sigma$ is a tuple $\cF = (\bar{\sigma}, I,
  \theta)$, where $\bar{\sigma} = (\sigma_1, \ldots, \sigma_N)$ is a
  list of pairwise distinct star-types over $\Sigma$, $I$ is a subset
  of \/ $\cI$, and $\theta$ is a function $\theta: I \rightarrow \Xi$
  such that, for all $\{i, i'\} \in I$ with $i \leq i'$, ${\rm
  tp}_1(\theta(\{i,i'\})) = \pi_i$ and ${\rm tp}_2(\theta(\{i,i'\})) =
  \pi_{i'}$.  The {\em dimension} of $\bar{\sigma}$ is $N$. For $Y$ a
  positive integer, $\cF$ is $Y$-{\em bounded} if, for all $k$ ($1
  \leq k \leq N$) and all $j$ ($1 \leq j \leq M$), $\sigma_k[j] \leq
  Y$. Finally, $\cF$ is {\em chromatic} if every $\sigma_k$ is
  chromatic.
\label{def:frame}
\end{definition}
Think of a frame $\cF$ as a (putative) statistical summary of a
structure $\fA$, specifically:
\begin{definition}
  Let $\cF = (\bar{\sigma}, I, \theta)$ be a frame over $\Sigma$, and
  let $\fA$ be a structure interpreting $\Sigma$. We say that $\cF$
  {\em describes} $\fA$ if the following conditions hold:
\begin{enumerate}
\item $\bar{\sigma}$ is a list of all and only those 
star-types realized in $\fA$;
\item $I$ is the set of all and only those $\{ i, i' \}$ ($1 \leq i
  \leq i' \leq L$) such that $\pi_i$ and $\pi_{i'}$ do not form a noisy
  pair in $\fA$ (see Definition~\ref{def:fused});
\item for each $\{ i, i' \} \in I$, there exist $a \in A_\pi$ and $a'
  \in A_{\pi'}$ such that $a \neq a'$ and ${\rm tp}^\fA[a,a'] =
  \theta(\{ i, i' \})$.
\end{enumerate}
\label{def:framefor}
\end{definition}

Every finitely branching structure $\fA$ is evidently described by
some (not necessarily unique) frame; and certain interesting
properties of $\fA$ correspond to obvious properties of the frames
which describe it, as we see from the following two lemmas.
\begin{lemma}
  Let $\fA$ be a finitely branching structure interpreting $\Sigma$
  and let $\cF$ be a frame over $\Sigma$ which describes $\fA$. Then:
  $(i)$ $\fA$ is chromatic if and only if $\cF$ is chromatic; $(ii)$
  $\fA$ is $X$-sparse if and only if $\cF$ has dimension at most $X$;
  and $(iii)$ if $\fA$ is $Y$-branching then $\cF$ has bound $Y$.
\label{lma:reflectedproperties}
\end{lemma}
\begin{proof}
Immediate.
\end{proof}
For the next definition, recall that a 1-type $\pi$ is simply a finite
collection of formulas, so that $\bigwedge \pi$ denotes the
conjunction of those formulas; similarly for 2-types.
\begin{definition}
Let $\cF = (\bar{\sigma}, I, \theta)$ be a frame over $\Sigma$, where
$\bar{\sigma} = (\sigma_1, \ldots, \sigma_N)$. We write $\cF
\models \phi^*$ if the following conditions are satisfied:
\begin{enumerate}
\item for all $k$ ($1 \leq k \leq N$) 
  $\models \bigwedge {\rm tp}(\sigma_k) \rightarrow \alpha$;
\label{item:fladescription0}
\item for all $k$ ($1 \leq k \leq N$) and all $j$ ($1 \leq j \leq M$),
  if $\sigma_k[j] > 0$ then \linebreak
  $\models \bigwedge \mu_j \rightarrow
  \beta(x,y) \wedge \beta(y,x)$;
\label{item:fladescription1}
\item for all $\{i,i'\} \in I$, $\models \bigwedge \theta(\{i,i'\})
  \rightarrow \beta(x,y) \wedge \beta(y,x)$;
\label{item:fladescription2}
\item for all $k$ ($1 \leq k \leq N$) and all $h$ ($1 \leq h \leq m$),
  the integers in the set $\{ \sigma_k[j] \mid 1 \leq j \leq M \mbox{
  and }  f_h(x,y) \in \mu_j \}$ sum to $C_h$.
\label{item:fladescription3}
\end{enumerate}
\label{def:fladescription}
\end{definition}
\begin{remark}
  Let $\cF= (\bar{\sigma}, I, \theta)$ be a frame over $\Sigma$, where
  $\bar{\sigma} = (\sigma_1, \ldots, \sigma_N)$. If $\cF \models
  \phi^*$, then, for all $k$ ($1 \leq k \leq N$), $\sum_{1 \leq j \leq
  M} \sigma_k[j] \leq mC$.
\label{remark:bound}
\end{remark}
\begin{lemma}
  Let $\fA$ be a finitely branching structure interpreting $\Sigma$
  and let $\cF$ be a frame over $\Sigma$ describing $\fA$. If $\fA
  \models \phi^*$, then $\cF \models \phi^*$.
\label{lma:fladescription}
\end{lemma} 
\begin{proof}
Almost immediate.
\end{proof}
However, while every finitely branching structure is described by some
frame, not every frame describes a structure; and it is important for
us to define a class of frames which do. Recall that $L$ is the number
of 1-types over $\Sigma$ and $M^*$ the number of invertible
message-types over $\Sigma$.
\begin{notation}
  Let $\cF = (\bar{\sigma}, I, \theta)$ be a frame over $\Sigma$,
  where $\bar{\sigma} = (\sigma_1, \ldots, \sigma_N)$. If $\cF$ is
  clear from context, for integers $i,j,k$ in the ranges $1 \leq i
  \leq L$, $1 \leq j \leq M^*$, $1 \leq k \leq N$ write:
\begin{align*}
o_{ik} & =
\begin{cases}
1 \mbox{ if ${\rm tp}(\sigma_k) = \pi_i$}\\
0 \mbox{ otherwise;}
\end{cases}\\
p_{ik} &= 
\begin{cases}
1 \mbox{ if, for all $j$ ($1 \leq j \leq M$), 
           ${\rm tp}_2(\mu_j) = \pi_i$ implies $\sigma_k[j]= 0$}\\
0 \mbox{ otherwise;}
\end{cases}\\
q_{jk} &= \sigma_k[j];\\
r_{ik} &= \sum \{ \sigma_k[j] \mid M^*+ 1 \leq j \leq M \mbox{ and } 
                                   {\rm tp}_2(\mu_j) = \pi_i \};\\
s_{ik} &= \sum \{ \sigma_k[j] \mid 1 \leq j \leq M \mbox{ and } 
                                   {\rm tp}_2(\mu_j) = \pi_i \}.
\end{align*}
\label{notation:description}
\end{notation}
\begin{remark}
Let $\cF$ be a frame over $\Sigma$, let $\fA$ be a structure interpreting
$\Sigma$, and suppose $\cF$ describes $\fA$. In that case, the symbols
$o_{ik}$, $p_{ik}$, $q_{jk}$, $r_{ik}$ and $s_{ik}$ in
Notation~\ref{notation:description} have the following interpretations
with respect to $\fA$:
\begin{enumerate}
\item $o_{ik} = 1$ just in case every element with star-type $\sigma_k$
  has 1-type $\pi_i$;
\item $p_{ik} = 1$ just in case no element with star-type $\sigma_k$
  sends a message to any element having 1-type $\pi_i$;
\item $q_{jk}$ counts how many messages of type $\mu_j$ any element
  having star-type $\sigma_k$ sends;
\item $r_{ik}$ is the total number elements of 1-type $\pi_i$ to which
  any element having star-type $\sigma_k$ sends a non-invertible
  message; and
\item $s_{ik}$ is the total number elements of 1-type $\pi_i$ to which
  any element having star-type $\sigma_k$ sends a message.
\end{enumerate}
\label{remark:framenotation}
\end{remark}
With this notation in hand we can characterize a class of frames whose
members are guaranteed to describe finite structures.
\begin{definition}
  Let $\cF = (\bar{\sigma}, I, \theta)$ be a frame over $\Sigma$,
  where $\bar{\sigma} = (\sigma_1, \ldots, \sigma_N)$. Let $\bar{w} =
  (w_1, \ldots, w_N)$ be a vector of positive integers.  Using
  Notation~\ref{notation:description}, for all $i$ ($1 \leq i \leq
  L$), all $i'$ ($1 \leq i' \leq L$) and all $j$ ($1 \leq j \leq
  M^*$), let:
\begin{equation*}
u_i = \sum_{1 \leq k \leq N} o_{ik} w_{k} \hspace{1cm}
v_j = \sum_{1 \leq k \leq N} q_{jk} w_{k} \hspace{1cm}
x_{ii'} = \sum_{1 \leq k \leq N} o_{ik}p_{i'k} w_{k}.
\end{equation*}
  Finally, let $Z$ be a positive integer. We say that $\bar{w}$ is a
  $Z$-{\em solution} of $\cF$ if the following conditions are
  satisfied for all $i$ ($1 \leq i \leq L$), all $i'$ ($1 \leq i' \leq
  L$), all $j$ ($1 \leq j \leq M^*$) and all $k$ ($1 \leq k \leq N$):
\begin{description}
\item[(C1) ] $v_j = v_{j'}$, where $j'$ is such that $\mu_j^{-1}
  = \mu_{j'}$;
\item[(C2) ] $s_{ik} \leq u_i$;
\item[(C3) ] $u_i \leq 1$ or $u_i >Z$;
\item[(C4) ] if $o_{ik} =1$, then either $u_i > 1$ or $r_{i'k} \leq
  x_{i'i}$;
\item[(C5) ] if $\{ i, i' \} \not \in I$, then either $u_i \leq 1$ or
  $u_{i'} \leq 1$;
\item[(C6) ] if $\{ i, i' \} \not \in I$ and $o_{ik} =1$, then either
  $u_i > 1$ or $r_{i'k} \geq x_{i'i}$.
\end{description}
\label{def:solvable}
\end{definition}
\begin{remark}
  Let $\cF$ be as in Definition~\ref{def:solvable}, and suppose that
  $\fA$ is a finite structure interpreting $\Sigma$ such that $\fA$ is
  described by $\cF$.  For all $k$ ($1 \leq k \leq N$), let $w_k$ be
  the number of elements of $A$ having star-type $\sigma_k$ in $\fA$.
  In that case, the symbols $u_i$, $v_j$ and $x_{ii'}$ in
  Definition~\ref{def:solvable} have the following interpretations
  with respect to $\fA$:
\begin{enumerate}
\item $u_i$ is the number of elements of $a \in A$ such that
  ${\rm tp}^\fA[a] = \pi_i$;
\item $v_j$ is the number of pairs $\langle a,b \rangle \in A^2$
  such that $a \neq b$ and ${\rm tp}^\fA[a,b] = \mu_j$;
\item $x_{ii'}$ is the number of elements of $a \in A$ such that ${\rm
tp}^\fA[a] = \pi_i$ and $a$ does not send a message to any element
having 1-type $\pi_{i'}$.
\end{enumerate}
\label{remark:interpretation}
\end{remark}

The following Lemma shows that our definition of $Z$-correctness is
not too stringent for the sorts of structures that interest us.
\begin{lemma}
  Let $\fA$ be a finite, $Y$-branching and
  $Z$-differentiated structure interpreting $\Sigma$, with
  $Z \geq (mY+1)^2$.  Let $\cF = (\bar{\sigma}, I, \theta)$ be a frame
  over $\Sigma$. If $\cF$ describes $\fA$, then $\cF$ is has a
  $Z$-solution.
\label{lma:maineasy}
\end{lemma}
\begin{proof}
Let $\bar{\sigma} = (\sigma_1, \ldots, \sigma_N)$, and let $w_k = |\{
a \in A : {\rm st}^\fA[a] = \sigma_k \}|$ for all $k$ ($1 \leq k \leq
N$).  We show that $\bar{w} = (w_1, \ldots, w_N)$ is a $Z$-solution of
$\cF$. In doing so, we make free use of
Remarks~\ref{remark:framenotation} and~\ref{remark:interpretation}.
Note that, by construction, the $w_1, \ldots, w_N$ are all positive.

\vspace{0.25cm}

\noindent
{\bf C1}: If $\mu_j^{-1} = \mu_{j'}$, then the sets $\{ \langle a, b
\rangle \mid a \neq b \mbox{ and } {\rm tp}^\fA[a,b] = \mu_j \}$ and
$\{ \langle a, b \rangle \mid a \neq b \mbox{ and } {\rm tp}^\fA[a,b]
= \mu_{j'} \}$ can obviously be put in 1--1 correspondence, namely:
$\langle a, b \rangle \mapsto \langle b, a \rangle$.  But the
cardinalities of these sets are $v_j$ and $v_{j'}$, respectively.

\vspace{0.25cm}

\noindent
{\bf C2}: Since $\cF$ describes $\fA$, any element of $A$ having
star-type $\sigma_k$ sends a message to exactly $s_{ik}$ elements having
1-type $\pi_i$. But $u_i$ is the number of elements of $A$ having
1-type $\pi_i$.  Since $\sigma_k$ is realized in $\fA$, $s_{ik} \leq
u_i$.

\vspace{0.25cm}

\noindent
{\bf C3}: Immediate given that $\fA$ is $Z$-differentiated.

\vspace{0.25cm}

\noindent
{\bf C4}: If $o_{ik} =1$ and $u_i \leq 1$, then $u_i = 1$, so that
$\fA$ contains exactly one element with 1-type $\pi_i$; moreover, this
element has star-type $\sigma_k$.  Denote this element by $a$. Thus, $a$
sends a non-invertible message to exactly $r_{i'k}$ elements 
with 1-type
$\pi_{i'}$. Clearly, none of these elements sends a message back to
$a$ (since otherwise $a$'s message to it would be invertible), so that
there exist at least $r_{i'k}$ elements with 1-type $\pi_{i'}$ which do
not send a message to $a$. But since $a$ is the only element with
1-type $\pi_i$, there exist at least $r_{i'k}$ elements with 1-type
$\pi_{i'}$ which do not send a message to any element of 1-type
$\pi_i$.  In other words, $r_{i'k} \leq x_{i'i}$.

\vspace{0.25cm}

\noindent
{\bf C5}: Since $\cF$ describes $\fA$, $\{i, i' \} \not \in I$ implies
that the 1-types $\pi_i$ and $\pi_{i'}$ form a noisy pair in $\fA$. In
that case, by Lemma~\ref{lma:fusion}, either $u_i \leq (mY+1)^2$ or
$u_{i'} \leq (mY+1)^2$.  Since $\fA$ is $Z$-differentiated and
$(mY+1)^2 \leq Z$, $u_i \leq 1$ or $u_{i'} \leq 1$.

\vspace{0.25cm}

\noindent
{\bf C6}: As already observed, if $o_{ik} =1$ and $u_i \leq 1$, then
$\fA$ contains exactly one element $a$ having 1-type $\pi_i$;
moreover, this element has star-type $\sigma_k$, and $x_{i'i}$ is the
number of elements of 1-type $\pi_{i'}$ which do not send a message to
$a$.  Since $\cF$ describes $\fA$, $\{i,i'\} \not \in I$ implies that
$\pi_i$ and $\pi_{i'}$ form a noisy pair, whence $a$ sends a
message---in fact, a non-invertible message---to all $x_{i'i}$ of
these elements. But since $a$ has star-type $\sigma_k$, $a$ sends a
non-invertible message to exactly $r_{i'k}$ elements having 1-type
$\pi_{i'}$.  Thus, $r_{i'k} \geq x_{i'i}$.
\end{proof}

We now prove a converse of Lemma~\ref{lma:maineasy}.
\begin{lemma}
  Let $\cF$ be a chromatic frame over $\Sigma$, and let $Z \geq
  3mC-1$. If $\cF$ has a $Z$-solution and $\cF \models \phi^*$, then
  there exists a finite structure $\fA$ interpreting $\Sigma$ such
  that $\fA \models \phi^*$.
\label{lma:mainhard}
\end{lemma}
\begin{proof}
  Let $\cF = (\bar{\sigma}, I, \theta)$, let $\bar{\sigma} =
  (\sigma_1, \ldots, \sigma_N)$, and let $\bar{w} = (w_1, \ldots,
  w_N)$ be a $Z$-solution of $\cF$.  In the sequel, we use the symbols
  $o_{ik}$, $p_{ik}$, $q_{jk}$, $r_{ik}$ and $s_{ik}$ (with indices in
  the appropriate ranges), as specified in
  Notation~\ref{notation:description}, and the symbols $u_i$, $v_j$
  and $x_{ii'}$ (again, with indices in the appropriate ranges), as
  specified in Definition~\ref{def:solvable}. Hence, the conditions
  {\bf C1}--{\bf C6} of Definition~\ref{def:solvable} hold.

  \vspace{0.25cm}

\noindent
  For every $k$ ($1 \leq k \leq N$), let $A_k$ be a set of cardinality
  $w_k$, and let $A$ be the disjoint union of the $A_k$. Think of
  $A_k$ as the set of elements which `want' to have star-type
  $\sigma_k$.  In addition, we define for all $i$ ($1 \leq i \leq L$),
  all $i'$ ($1 \leq i' \leq L$) and all $j$ ($1 \leq j \leq M^*$):
\begin{align*}
U_i =& \bigcup \{A_k \mid 1 \leq k \leq N \mbox{ and } o_{ik} = 1 \} \\ 
X_{ii'} =& 
  \bigcup \{A_k \mid 1 \leq k \leq N \mbox{ and } o_{ik}p_{i'k} = 1 \}\\
V_j =& \bigcup \{A_k \mid 1 \leq k \leq N \mbox{ and } q_{jk} = 1 \}.
\end{align*}
Since $\cF$ is chromatic, $q_{jk} \leq 1$ for all $j$ ($1 \leq j \leq
M^*$) and all $k$ ($1 \leq k \leq N$).  Thus, for all $i$, $i'$ and
$j$ in the appropriate ranges:
\begin{equation*}
|U_i| = u_i; \hspace{1cm}
|X_{ii'}| = x_{ii'}; \hspace{1cm}
|V_j| = v_j.
\end{equation*}
Think of $U_i$ as the set of elements which `want' to have 1-type
$\pi_i$, $X_{ii'}$ as the set of elements in $U_i$ which do not `want'
to send a message to any element in $U_{i'}$, and $V_j$ as the set of
elements which `want' to send an (invertible) message of type $\mu_j$
to some other element. We remark that $A_k \subseteq U_i$ if and only
if ${\rm tp}(\sigma_k) = \pi_i$.  Moreover, for all $j$ ($1 \leq j
\leq M^*$), there exists a unique $i$ ($ 1 \leq i \leq L$) such that
$V_j \subseteq U_i$---namely, that $i$ such that ${\rm tp}_1(\mu_j) =
\pi_i$. We now convert the domain $A$ into a structure $\fA$ in four
steps.

\vspace{0.25cm}

\noindent
{\bf Step 1} (Interpreting the unary predicates and diagonals of
binary predicates): For every $k$ ($1 \leq k \leq N$) and every $a \in
A_k$, set ${\rm tp}^\fA[a] = {\rm tp}(\sigma_k)$. At the end of this
step, we have, for every $i$ ($1 \leq i \leq L$) and every $a \in
U_i$, ${\rm tp}^\fA[a] = \pi_i$.

\vspace{0.25cm}

\noindent
{\bf Step 2} (Fixing the invertible message-types): For every $j$ ($1
\leq j \leq M^*$), let $j'$ be such that $\mu_j^{-1} = \mu_{j'}$.  By
{\bf C1}, $V_j$ and $V_{j'}$ are equinumerous.  If
$j' >j$, pick some 1--1 correspondence between $V_j$ and $V_{j'}$; and
for every $a \in V_j$, set ${\rm tp}^\fA[a,a'] = \mu_j$, where $a'$ is
the element of $V_{j'}$ corresponding to $a \in V_j$. This completes
Step~2.  We must show that these assignments are meaningful, do not
clash with Step~1, and do not clash with each other. Suppose then
that the assignment ${\rm tp}^\fA[a,a'] = \mu_j$ is made, and that
$\mu_j^{-1} = \mu_{j'}$.  Thus, $a \in V_j$ and $a' \in V_{j'}$. To
show that the assignment is meaningful, we must prove that $a \neq
a'$.  For contradiction, suppose $a = a'$, and let $k$ be such that $a
\in A_k$.  But then $\sigma_k[j]>0$ and $\sigma_k[j']>0$, which is
impossible by Remark~\ref{remark:chromaticprofilevector}.  To show
that the assignment does not clash with Step~1, suppose $\mu_j^{-1} =
\mu_{j'}$, and let $i$, $i'$ be such that $V_j \subseteq U_i$ and
$V_{j'} \subseteq U_{i'}$. As observed above, $\pi_i = {\rm
  tp}_1(\mu_j)$ and $\pi_{i'} = {\rm tp}_1(\mu_{j'}) = {\rm
  tp}_2(\mu_j)$, which conforms to the assignments in Step 1.  To show
that these assignments do not clash with each other, it suffices to
prove that, if $a \in V_j \cap V_h$, $a' \in V_{j'} \cap V_{h'}$
$\mu_j^{-1} = \mu_{j'}$ and $\mu_h^{-1} = \mu_{h'}$, then $j = h$.
Suppose then that antecedent of this conditional holds; let $k$ and
$k'$ be such that $a \in A_k$ and $a' \in A_{k'}$. Then
$\sigma_{k'}[j']>0$ and $\sigma_{k'}[h']>0$.  Since $\sigma_{k'}$ is a
star-type, ${\rm tp}_1(\mu_{j'})={\rm tp}_1(\mu_{h'})$, whence ${\rm
  tp}_2(\mu_j)={\rm tp}_2(\mu_h)$. But $\sigma_{k}[j]>0$ and
$\sigma_{k}[h]>0$, and since $\sigma_k$ is a {\em chromatic}
star-type, $j = h$.  Note that, if $\mu_j^{-1} = \mu_j$, then $V_j =
\emptyset$ by Remark~\ref{remark:chromaticprofilevector}. Thus, at the
end of Step 2, for every element $a \in A$ and and every $j$ ($1 \leq
j \leq M^*$), $a$ sends a (unique) message of type $\mu_j$ to some
other element if and only if $a \in V_j$.  That is: for all $k$ ($1
\leq k \leq N$), all $a \in A_k$, and all $j$ ($1 \leq j \leq M^*$),
there are exactly $\sigma_k[j]$ elements $a' \in A$ such that $a \neq
a'$ and ${\rm tp}^\fA[a,a']$ has been set to the (invertible)
message-type $\mu_j$.  We make one further observation before
proceeding. Suppose that ${\rm tp}^\fA[a,a']$ is assigned in this step
and that $a \in U_i$; we claim that $a' \not \in X_{i'i}$ for any
$i'$. To see this, suppose $a \in A_{k} \subseteq V_{j}$ and $a' \in
A_{k'} \subseteq V_{j'}$, with $\mu_j^{-1} = \mu_{j'}$. If $a \in
U_i$, then ${\rm tp}_1(\mu_j) = {\rm tp}_2(\mu_{j'}) = \pi_i$. But
$\sigma_{k'}[j] >0$, whence $p_{ik'} = 0$, whence $a' \not \in
X_{i'i}$. This observation will be useful in Step~3.

\vspace{0.25cm}

\noindent
{\bf Step 3} (Fixing the non-invertible message-types): Let $i$ and
$i'$ be such that $1 \leq i \leq i' \leq L$. We fix all the
non-invertible messages sent, in either direction, between $U_i$ and
$U_{i'}$. By {\bf C3}, either $u_i \leq 1$ or $u_i > Z$; similarly,
either $u_{i'} \leq 1$ or $u_{i'} > Z$.  We consider five cases.
\begin{figure}
\begin{center}
\input{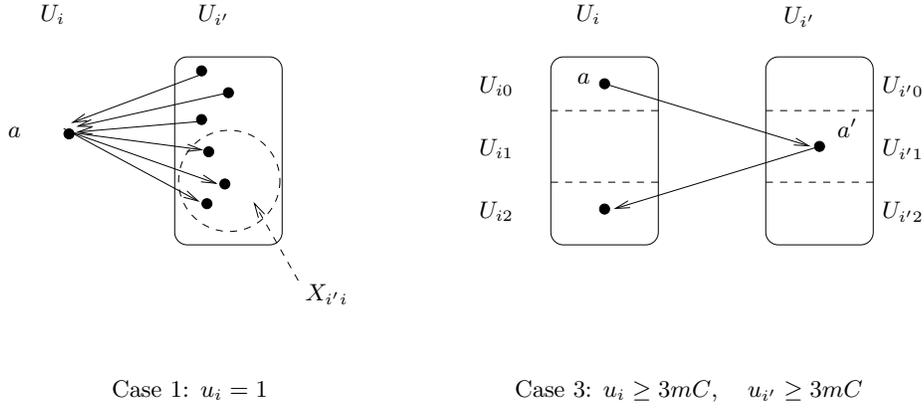}
\end{center}
\caption{Dealing with non-invertible messages.}
\label{fig:cases}
\end{figure}

\vspace{0.25cm}

\noindent
{\em Case~1: } $u_i =0$. In this case, there are no elements of $U_i$
and hence no 2-type assignments to be made between elements of $U_i$
and elements of $U_{i'}$.  Note that, by {\bf C2}, $s_{ik} = 0$ for
all $k$ ($1 \leq k \leq N$), whence $\sigma_k[j] = 0$ for all $k$ ($1
\leq k \leq N$) and for all $j$ ($1 \leq j \leq M$) such that ${\rm
tp}_2(\mu_j) = \pi_i$. (Intuitively, no element of $A$---and in
particular of $U_{i'}$---`wants' to send a message to an element with
1-type $\pi_i$ anyway.)

\vspace{0.25cm}

\noindent
{\em Case~2: } $u_i = 1$.  Let $a$ be the sole element of $U_i$, and
let $k$ be such that $a \in A_k$.  We deal first with the assignment
of non-invertible messages sent from $U_{i'}$ to $U_i = \{ a \}$.
Consider any $a' \in A_{k'} \subseteq U_{i'}$.  By~{\bf C2}, $s_{ik'}
\leq 1$; hence there is at most one value of $j$ in the range $1 \leq
j \leq M$ such that ${\rm tp}_2(\mu_j) = \pi_i$ and $\sigma_{k'}[j]
>0$. Suppose then that such a $j$ exists.  Again, since $s_{ik'} \leq
1$, $\sigma_{k'}[j] =1$. If $j \leq M^*$, then this message has
already been dealt with in Step 2; otherwise, set ${\rm tp}^\fA[a',a]
= \mu_j$. Since ${\rm tp}_1(\mu_j) = \pi_{i'}$ and ${\rm tp}_2(\mu_j)
= \pi_{i}$, this assignment does not clash with Step~1.  Observe also
that, just as in Step~2, if this assignment is made, we have, by
definition, $p_{ik'} = 0$, so that $a' \not \in X_{i'i}$.  By carrying
out the same procedure for all $a' \in U_{i'}$, we complete the
assignment of non-invertible messages sent from $U_{i'}$ to $U_i$.  It
remains to deal with the non-invertible messages sent from $U_i = \{ a
\}$ to $U_{i'}$.  Remembering that $a \in A_k$, {\bf C4} ensures the
existence of a subset $R \subseteq X_{i'i}$ such that $|R| = r_{i'k}$.
For each $j$ ($M^*+1 \leq j \leq M$), if ${\rm tp}_2(\mu_j) =
\pi_{i'}$, select $q_{jk}$ fresh elements $a'$ of $R$, and make the
assignment ${\rm tp}^\fA[a,a'] = \mu_j$.  (There are enough such
elements by the definition of $r_{i'k}$.)  Since ${\rm tp}_1(\mu_j) =
\pi_i$ and ${\rm tp}_2(\mu_j) = \pi_{i'}$, this assignment does not
clash with Step~1.  Moreover, we have observed that ${\rm
  tp}^\fA[a,a']$ has previously been assigned (either in this step or
in Step~2) only if $a' \not \in X_{i'i}$.  Thus, these assignments do
not clash with those made earlier in this step or those made in
Step~2.  The situation is illustrated in Fig.~\ref{fig:cases} Case~2.

\vspace{0.25cm}

\noindent
{\em Case~3: } $u_{i'} =0$ and $u_i > Z$.  Symmetrical to Case 1.

\vspace{0.25cm}

\noindent
{\em Case~4: } $u_{i'} =1$ and $u_i > Z$.  Symmetrical to Case 2.

\vspace{0.25cm}

\noindent
{\em Case~5: } $u_i > Z$ and $u_{i'} > Z$.  Since $Z \geq 3mC-1$,
partition $U_i$ into three sets $U_{i0}$, $U_{i1}$, $U_{i2}$, each
containing at least $mC$ elements; and similarly for $U_{i'}$.
Suppose $a \in U_{i}$.  Then for some $h$ ($0 \leq h < 3$), $a \in
U_{ih}$.  Let $k$ be such that $a \in A_k$, and let $h' = h+1$ (mod
3). For all $j$ ($M^*+1 \leq j \leq M$), pick $\sigma_k[j]$ fresh
elements $a'$ of $U_{i'h'}$ such that ${\rm tp}^\fA[a,a']$ was not
assigned in Step~2, and set ${\rm tp}^\fA[a,a'] = \mu_j$.  By
Remark~\ref{remark:bound}, we have, for each $k$ ($1 \leq k \leq N$),
$\sum_{1 \leq j \leq M} \sigma_k[j] \leq mC$; and since $|U_{i'h'}|
\geq mC$, we never run out of fresh elements to pick. In this way, we
deal with all messages sent from $U_i$ to $U_{i'}$; the messages sent
from $U_{i'}$ to $U_i$ are dealt with symmetrically.  It is obvious
that these assignments do not clash with Step~1 (or Step~2); and the
fact that $h' = h+1$ (mod 3), ensures that they do not clash with each
other (even if $i = i'$), as is evident from Fig.~\ref{fig:cases}
Case~5.

\vspace{0.25cm}

\noindent
Performing these assignments for all pairs $i, i'$ such that $1 \leq i
\leq i' \leq L$ completes Step~3.  At the end of Step 3, then, for all
$k$ ($1 \leq k \leq N$), all $a \in A_k$, and all $j$ ($1 \leq j \leq
M$), there are exactly $\sigma_k[j]$ elements $a' \in A$ such that $a
\neq a'$ and ${\rm tp}^\fA[a,a']= \mu_j$.

\vspace{0.25cm}

\noindent
{\bf Step 4} (Fixing the silent 2-types): Finally, we deal with pairs
of distinct elements $a,a'$ whose 2-type in $\fA$ has not been yet
been assigned. Let $i, i'$ be such that $a \in U_i$ and $a' \in
U_{i'}$, and suppose, without loss of generality, that $i \leq i'$.
We claim that $\langle i, i' \rangle \in I$. For suppose otherwise. By
{\bf C5}, we have either $u_i = 1$ or $u_{i'} = 1$. Assume the
former. Now let $k, k'$ be such that $a \in A_k$ and $a' \in A_{k'}$.
Thus, $o_{i'k'} = 1$. If $p_{ik'} = 0$ then there is some $j$ ($1 \leq
j \leq M$) such that $q _{jk'} = \sigma_{k'}[j] > 0$ and ${\rm
tp}_2(\mu_j) = \pi_i$, whence---bearing in mind that $a$ is the unique
element of $U_i$---${\rm tp}^\fA[a,a']$ will certainly have been
assigned in Step~2 (if $\mu_j$ is an invertible message-type) or in
Step~3 Case~2 (if $\mu_j$ is a non-invertible message-type),
contradicting our supposition. So we may assume that $p_{ik'} = 1$ and
hence $o_{i'k'}p_{ik'} = 1$. That is: $a \in X_{i'i}$. But $|X_{i'i}|
= x_{i'i}$.  And by~{\bf C6}, $x_{i'i} \leq r_{i'k}$. Yet in Step~3
(Case~2), $r_{i'k}$ elements of $X_{i'i}$ were chosen to receive
messages from $a$.  Hence $a'$ must be among these elements, so that
${\rm tp}^\fA[a,a']$ has already been assigned, again a contradiction.
The case where $u_{i'} \leq 1$ proceeds symmetrically.  Thus, we have
established that, if ${\rm tp}^\fA[a,a']$ has not yet been assigned,
then $\{ i, i' \} \in I$, so that we can make the assignment ${\rm
tp}^\fA[a,a'] = \theta(\{i,i'\})$.  Since ${\rm
tp}_1(\theta(\{i,i'\})) = \pi_i$ and ${\rm tp}_2(\theta(\{i,i'\})) =
\pi_{i'}$, there is no clash with Step~1. Evidently, we can proceed in
this way until all remaining 2-types have been assigned.

\vspace{0.25cm}

\noindent
This completes the construction of $\fA$.  The only 1-types realized
in $\fA$ are the 1-types ${\rm tp}(\sigma_k)$ (where $1 \leq k \leq
N$).  The only message-types realized in $\fA$ are those $\mu_j$ such
that $\sigma_k[j] >0$ for some $k$. And the only silent 2-types
realized in $\fA$ are the $\theta(\{i,i'\})$ for $\{i,i'\} \in I$.
Since $\cF \models \phi^*$, we have $\fA \models \forall x \alpha
\wedge \forall x \forall y (\beta \vee x \approx y)$.  Moreover, for
all $k$ ($1 \leq k \leq N$), and for all $a \in A_k$, we have ${\rm
  st}^\fA[a] = \sigma_k$.  Since $\cF \models \phi^*$, we have, for
all $a \in A$, ${\rm ct}^\fA[a] = (C_1, \ldots, C_m)$; in other words,
$\fA \models \bigwedge_{1 \leq h \leq m} \forall x \exists_{=C_h} y
(f_h(x,y) \wedge x \not \approx y)$. Hence, $\fA \models \phi^*$.
\end{proof}
We can now employ a standard result to bound the complexity of
determining whether a given frame has a $Z$-solution.
\begin{lemma}
Let $\cF$ be a $Y$-bounded, $N$-dimensional frame over $\Sigma$, and
let $Z$ be an integer. Then $\cF$ has a $Z$-solution if and only if it
has a $Z$-solution $\bar{w}$ such that every component of $\bar{w}$ is
bounded by some (fixed) singly exponential function of the quantity
$L+M^*+N+\log Y + \log Z$.
\label{lma:small}
\end{lemma}
\begin{proof}
  This follows immediately from the well-known result
  (Papadimitriou~\cite{logic:papadimitriou81}) that, if an integer
  programming problem has a solution at all, then it has a solution
  all of whose components are bounded by a singly exponential function
  of the size of the problem (encoded in the obvious way). But the
  conditions {\bf C1}--{\bf C6} in Definition~\ref{def:solvable}
  amount to a disjunction of integer programming problems whose sizes
  are all bounded by a polynomial function of $L+M^*+N+\log Y + \log
  Z$.
\end{proof}

\begin{theorem}
The problem Fin-Sat-$\cC^2$ is in NEXPTIME.
\label{theo:finsat}
\end{theorem}
\begin{proof}
Given $\phi$, let $m$, $f_1, \ldots, f_m$, $C_1, \ldots, C_m$ and
$\phi^*$ be as in Lemma~\ref{lma:normalform}, and let $C = \max_{1
\leq h \leq m} C_h$; thus $C \leq 2^{\lVert \phi^* \rVert}$. Let $Z =
\max(3mC-1,(mC+1)^2)$. Let $\Sigma$ be the signature of $\phi^*$
together with $\log((mC)^2+1) + \log Z$ (rounded up) additional
unary predicates, regarded as a counting signature by taking the
counting predicates of $\Sigma$ to be $f_1, \ldots, f_m$.  Denote the
total size of $\Sigma$ by $s$: evidently, $s$ is bounded by a
polynomial function of $\lVert \phi^* \rVert$.  Finally, let $X =
4^s.(16^s +1)(C+1)^{sm}$.

\vspace{0.25cm}

\noindent
Since $C \leq 2^{\lVert \phi^* \rVert}$, $C$ is bounded by a singly
exponential function of $\lVert \phi \rVert$. Thus, the problem of
determining the satisfiability of $\phi$ over structures of size $C$
or less is obviously in NEXPTIME. By Lemma~\ref{lma:normalform} then,
it suffices to show that the satisfiability of $\phi^*$ can be decided
nondeterministically in time bounded by a singly exponential function
of $\lVert \phi^* \rVert$.

\vspace{0.25cm}

\noindent
We claim that $\phi^*$ is finitely satisfiable if and only if there
exists a chromatic, $C$-bounded frame $\cF$ over $\Sigma$ of dimension
$N \leq X$, such that $\cF$ has a $Z$-solution and $\cF \models
\phi^*$.  For suppose $\phi^*$ is finitely satisfiable.  By
Lemmas~\ref{lma:chromatic} and~\ref{lma:differentiated}, $\phi^*$ has
a finite, chromatic, $Z$-differentiated model $\fA$ interpreting
$\Sigma$.  By Lemma~\ref{lma:sparse}, we may assume without loss of
generality that $\fA$ is also $X$-sparse. Let $\cF$ be a frame over
$\Sigma$ describing $\fA$. By Lemma~\ref{lma:reflectedproperties},
$\cF$ is chromatic, of dimension $N \leq X$ and $C$-bounded.  By
Lemma~\ref{lma:fladescription}, $\cF \models \phi^*$. Since $Z \geq
(mC+1)^2$, Lemma~\ref{lma:maineasy} guarantees that $\cF$ has a
$Z$-solution. Conversely, suppose $\cF$ is a chromatic frame over
$\Sigma$ such that $\cF$ has a $Z$-solution and $\cF \models
\phi^*$. Since $Z \geq 3mC-1$, Lemma~\ref{lma:mainhard} guarantees
that $\phi^*$ is finitely satisfiable.

\vspace{0.25cm}

\noindent
By Lemma~\ref{lma:small}, then, $\phi^*$ is satisfiable if and only if
there exists a chromatic, $C$-bounded frame $\cF$ over $\Sigma$ of
dimension $N \leq X$ and a vector $\bar{w}$ of positive integers
bounded by some doubly exponential function of $\lVert \phi^* \rVert$,
such that $\cF \models \phi^*$ and $\bar{w}$ is a $Z$-solution of
$\cF$. Using the standard binary encoding of integers, it is easy to
write down $\cF$ and $\bar{w}$ in a number of bits bounded by a singly
exponential function of $\lVert \phi^* \rVert$, and to check whether
they satisfy the above conditions in time bounded by a singly
exponential function of $\lVert \phi^* \rVert$.  This completes the
proof.
\end{proof}

We note in passing that the above proof yields a small model property
for finitely satisfiable $\cC^2$-formulas.
\begin{corollary}
Let $\phi$ be a formula of $\cC^2$. If $\phi$ is finitely satisfiable,
then it is satisfiable in a structure of size bounded by a doubly
exponential function of $\lVert \phi \rVert$.
\label{cor:smp}
\end{corollary}
\begin{proof}
The structure built in Lemma~\ref{lma:mainhard} from $\cF$ and its
$Z$-solution $\bar{w}$ has domain of cardinality $w_1 + \cdots + w_N$.
\end{proof}
Notice that the complexity result of Theorem~\ref{theo:finsat} is
better than one might na\"{\i}vely expect on the basis of the small
model property of Corollary~\ref{cor:smp}. Nevertheless, the bound of
Corollary~\ref{cor:smp} is optimal in the sense that there is a
sequence $\{\phi_i\}$ of finitely satisfiable formulas of $\cC^2$ whose
size grows as a polynomial function of $i$, but whose smallest
satisfying structures grow as a doubly exponential function of $i$
(Gr\"{a}del and Otto~\cite{logic:gor97}, p.~317).
\section{Deciding satisfiability}
Having established the complexity of determining finite satisfiability
in $\cC^2$, we now turn to the corresponding (general) satisfiability
problem. In fact, there is almost no further work to do.
\begin{notation}
  Let $\bbN^*$ denote the set $\bbN \cup \{ \aleph_0 \}$.  We extend
  the ordering $>$ and the arithmetic operations $+$ and $\cdot$ from
  $\bbN$ to $\bbN^*$ in the obvious way. Specifically, we define
  $\aleph_0 > n$ for all $n \in \bbN$; we define $\aleph_0 + \aleph_0
  = \aleph_0 \cdot \aleph_0 = \aleph_0$ and $0 \cdot \aleph_0 =
  \aleph_0 \cdot 0 = 0$; we define $n + \aleph_0 = \aleph_0 +n =
  \aleph_0$ for all $n \in \bbN$; and we define $n \cdot \aleph_0 =
  \aleph_0 \cdot n = \aleph_0$ for all $n \in \bbN$ such that $n >0$.
  Under this extension, $>$ remains a total order, and $+$, $\cdot$
  remain associative and commutative.
\label{notation:arithmetic}
\end{notation}
A system of linear equalities and inequalities defining an integer
programming problem can of course be re-interpreted so that solutions
are sought not over $\bbN$ but over $\bbN^*$. (We always assume that
the coefficients occurring in such problems are in $\bbN$.)  As an
example, the single inequality $x_1 \geq x_1 +1$ has no solutions over
$\bbN$, but it does have a solution over $\bbN^*$, namely, $x_1 =
\aleph_0$.

Lemmas~\ref{lma:chromatic}--\ref{lma:sparse} apply to both finite and
infinite structures. Furthermore, the definition of a frame and its
relationship to the (finitely branching) structures it describes make no
reference to the cardinalities of those structures, and
Lemmas~\ref{lma:reflectedproperties}--\ref{lma:fladescription} again
apply generally.  The concept of a $Z$-solution introduced in
Definition~\ref{def:solvable} requires extension, however.
\begin{definition}
  Let $\Sigma$ and $\cF = (\bar{\sigma}, I, \theta)$ be as in
  Definition~\ref{def:solvable}, and let $Z$ be a positive integer.
  Let $\bar{w} = (w_1, \ldots, w_N)$ be a vector of non-zero elements
  of $\bbN^*$.  We say that $\bar{w}$ is an {\em extended}
  $Z$-solution of $\cF$ if $\bar{w}$ satisfies the conditions of
  Definition~\ref{def:solvable}, with the arithmetic interpreted over
  $\bbN^*$ as specified in Notation~\ref{notation:arithmetic}.
\label{def:extendedsolution}
\end{definition}
We must check that the obvious analogues
of Lemmas~\ref{lma:maineasy} and~\ref{lma:mainhard} hold:
\begin{lemma}
  Let $\fA$ be a $Y$-branching and 
  $Z$-differentiated structure interpreting
  $\Sigma$, with $Z \geq (mY+1)^2$.
  Let $\cF = (\bar{\sigma}, I, \theta)$ be a frame over $\Sigma$. If
  $\cF$ describes $\fA$, then $\cF$ has an extended $Z$-solution.
\label{lma:maineasyextended}
\end{lemma}
\begin{lemma}
  Let $\cF$ be a chromatic frame over $\Sigma$, and let $Z \geq
  3mC-1$. If $\cF$ has an extended $Z$-solution and $\cF \models
  \phi^*$, then there exists a structure $\fA$ interpreting $\Sigma$
  such that $\fA \models \phi^*$.
\label{lma:mainhardextended}
\end{lemma}
The proofs are exactly the same as in the finite case. Note that the
variables $u_i$, $v_j$ and $x_{ii'}$ as well as the $w_k$ may now take
the value $\aleph_0$; by contrast, the coefficients $o_{ik}$,
$p_{ik}$, $q_{jk}$, $r_{ik}$ and $s_{ik}$ remain finite.
Remark~\ref{remark:interpretation} generalizes unproblematically to
countably infinite structures, so that the quantities $u_i$, $v_j$ and
$x_{ii'}$ mentioned in Definition~\ref{def:solvable} continue to have
their familiar interpretations. The proofs of
Lemmas~\ref{lma:maineasyextended} and~\ref{lma:mainhardextended} then
proceed exactly as before.

There is one final hurdle to overcome. The proof of
Lemma~\ref{lma:small} used a well-known result bounding solutions of
integer programming problems.  Since we are dealing with
$\bbN^*$-programming problems, we need the following extension of that
result.
\begin{lemma}
  Let $\Phi$ be a finite set of linear inequalities of the form
\[
a_0 + a_1 x_1 + \cdots + a_n x_n \leq
b_0 + b_1 x_1 + \cdots + b_n x_n
\]
in variables $x_1, \ldots, x_n$. Here, all coefficients are assumed to
be in $\bbN$.  We take the size of $\Phi$, denoted $\lVert \Phi
\rVert$, to be measured in the usual way, assuming binary encoding of
integers.  If $\Phi$ has a solution over $\bbN^*$, then $\Phi$ has a
solution over $\bbN^*$ such that all finite values are bounded by some
(fixed) singly exponential function of $\lVert \Phi \rVert $.
\label{lma:extend}
\end{lemma}
\begin{proof} Suppose that $\Phi$ has a solution over $\bbN^*$.
  Re-order the variables if necessary so that this solution has the
  form $\bar{a}\bar{\aleph}_0$, with $\bar{a} = a_1, \ldots, a_k \in
  \bbN^k$ for some $k$ ($0 \leq k \leq n$) and $\bar{\aleph}_0$ a
  $(n-k)$-tuple of $\aleph_0$s. Say that an inequality in $\Phi$ {\em
    does not involve} the variable $x_i$ if the corresponding
  coefficients $a_i$ and $b_i$ are both zero.  Let $\Psi$ be the set
  of inequalities in $\Phi$ involving none of the $x_{k+1}, \ldots,
  x_n$. Thus, $\Psi$, considered as a problem in variables $x_1,
  \ldots, x_k$, has a solution $\bar{a}$ over $\bbN$, whence it has
  has a solution $\bar{a}'$ bounded by some singly exponential
  function of $\lVert \Psi \rVert$ (and hence of $\lVert \Phi
  \rVert$). But then it is easy to see that $\bar{a}'\bar{\aleph}_0$
  is a solution of $\Phi$.
\end{proof}

\begin{theorem}
   The problem Sat-$\cC^2$ is in NEXPTIME.
\label{theo:sat}
\end{theorem}
\begin{proof}
   Exactly as for Theorem~\ref{theo:finsat}, noting that, by
   Lemma~\ref{lma:extend}, the existence of extended $Z$-solutions for
   a frame $\cF$ can be checked nondeterministically in time bounded
   by an exponential function of $\lVert \phi^* \rVert$.
\end{proof}
Obviously, there is no interesting small model property for
satisfiable $\cC^2$-formulas corresponding to Corollary\ref{cor:smp};
however, we have the next best thing.
\begin{corollary}
  Let $\phi$ be a formula of $\cC^2$.  Then there exist integers $X$
  and $W$, with $X$ bounded by a singly exponential function of
  $\lVert \phi \rVert$ and $W$ by a doubly exponential function of
  $\lVert \phi \rVert$ such that, if $\phi$ is satisfiable, then it is
  satisfiable in an $X$-sparse structure in which every star-type is
  realized either infinitely often or at most $W$ times.
\label{cor:extendedsmp}
\end{corollary}
\bibliographystyle{plain} 
\bibliography{logic}
\end{document}